\documentclass[twocolumn]{aastex63}
\pdfoutput=1 
\usepackage{amsmath,amstext,amssymb}
\usepackage[T1]{fontenc}
\usepackage{apjfonts} 
\usepackage{subfigure}
\usepackage{xcolor}
\usepackage{soul}
\usepackage{comment}
\usepackage{graphicx}
\usepackage[normalem]{ulem}
\setstcolor{red}

\graphicspath{{figures/}}

\newcommand{\de}{{\rm d}}
\newcommand{\phz}{photo--$z$~}

\shorttitle{Standard siren measurement using GW190814 and DES}
\shortauthors{A. Palmese et al.}

\begin{document}

\title{
A statistical standard siren measurement of the Hubble constant from the LIGO/Virgo gravitational wave compact object merger GW190814 and Dark Energy Survey galaxies}

\author{A.~Palmese}
\affiliation{Fermi National Accelerator Laboratory, P. O. Box 500, Batavia, IL 60510, USA}
\affiliation{Kavli Institute for Cosmological Physics, University of Chicago, Chicago, IL 60637, USA}
\correspondingauthor{Antonella Palmese}
\email{palmese@fnal.gov}

\author{J.~deVicente}
\affiliation{Centro de Investigaciones Energ\'eticas, Medioambientales y Tecnol\'ogicas (CIEMAT), Madrid, Spain}

\author{M.~E.~S.~Pereira}
\affiliation{Department of Physics, Brandeis University, Waltham, MA 02453, USA}

\author{J.~Annis}
\affiliation{Fermi National Accelerator Laboratory, P. O. Box 500, Batavia, IL 60510, USA}

\author{W.~Hartley}
\affiliation{D\'{e}partement de Physique Th\'{e}orique and Center for Astroparticle Physics, Universit\'{e} de Gen\`{e}ve, 24 quai Ernest Ansermet, CH-1211 Geneva, Switzerland}
\affiliation{Department of Physics \& Astronomy, University College London, Gower Street, London, WC1E 6BT, UK}
\affiliation{Department of Physics, ETH Zurich, Wolfgang-Pauli-Strasse 16, CH-8093 Zurich, Switzerland}

\author{K.~Herner}
\affiliation{Fermi National Accelerator Laboratory, P. O. Box 500, Batavia, IL 60510, USA}

\author{M.~Soares-Santos}
\affiliation{Department of Physics, Brandeis University, Waltham, MA 02453, USA}

\author{M.~Crocce}
\affiliation{Institut d'Estudis Espacials de Catalunya (IEEC), 08034 Barcelona, Spain}
\affiliation{Institute of Space Sciences (ICE, CSIC),  Campus UAB, Carrer de Can Magrans, s/n,  08193 Barcelona, Spain}

\author{D.~Huterer}
\affiliation{Department of Physics, University of Michigan, Ann Arbor, MI 48109, USA}

\author{I. Maga\~na~Hernandez}
\affiliation{University of Wisconsin-Milwaukee, Milwaukee, WI 53201, USA}

\author{A.~Garcia}
\affiliation{Department of Physics, Brandeis University, Waltham, MA 02453, USA}

\author{J.~Garcia-Bellido}
\affiliation{Instituto de Fisica Teorica UAM/CSIC, Universidad Autonoma de Madrid, 28049 Madrid, Spain}

\author{J.~Gschwend}
\affiliation{Laborat\'orio Interinstitucional de e-Astronomia - LIneA, Rua Gal. Jos\'e Cristino 77, Rio de Janeiro, RJ - 20921-400, Brazil}
\affiliation{Observat\'orio Nacional, Rua Gal. Jos\'e Cristino 77, Rio de Janeiro, RJ - 20921-400, Brazil}

\author{D.~E.~Holz}
\affil{Kavli Institute for Cosmological Physics, University of Chicago, Chicago, IL 60637, USA}

\author{R.~Kessler}
\affiliation{Department of Astronomy and Astrophysics, University of Chicago, Chicago, IL 60637, USA}
\affiliation{Kavli Institute for Cosmological Physics, University of Chicago, Chicago, IL 60637, USA}

\author{O.~Lahav}
\affiliation{Department of Physics \& Astronomy, University College London, Gower Street, London, WC1E 6BT, UK}

\author{R.~Morgan}
\affiliation{Physics Department, 2320 Chamberlin Hall, University of Wisconsin-Madison, 1150 University Avenue Madison, WI  53706-1390}

\author{C.~Nicolaou}
\affiliation{Department of Physics \& Astronomy, University College London, Gower Street, London, WC1E 6BT, UK}

\author{C.~Conselice}
\affiliation{University of Nottingham, School of Physics and Astronomy, Nottingham NG7 2RD, UK}

\author{R.~J.~Foley}
\affiliation{Santa Cruz Institute for Particle Physics, Santa Cruz, CA 95064, USA}

\author{M.~S.~S.~Gill}
\affiliation{SLAC National Accelerator Laboratory, Menlo Park, CA 94025, USA}

\author{T.~M.~C.~Abbott}
\affiliation{Cerro Tololo Inter-American Observatory, NSF's National Optical-Infrared Astronomy Research Laboratory, Casilla 603, La Serena, Chile}
\author{M.~Aguena}
\affiliation{Departamento de F\'isica Matem\'atica, Instituto de F\'isica, Universidade de S\~ao Paulo, CP 66318, S\~ao Paulo, SP, 05314-970, Brazil}
\affiliation{Laborat\'orio Interinstitucional de e-Astronomia - LIneA, Rua Gal. Jos\'e Cristino 77, Rio de Janeiro, RJ - 20921-400, Brazil}
\author{S.~Allam}
\affiliation{Fermi National Accelerator Laboratory, P. O. Box 500, Batavia, IL 60510, USA}
\author{S.~Avila}
\affiliation{Instituto de Fisica Teorica UAM/CSIC, Universidad Autonoma de Madrid, 28049 Madrid, Spain}
\author{K.~Bechtol}
\affiliation{Physics Department, 2320 Chamberlin Hall, University of Wisconsin-Madison, 1150 University Avenue Madison, WI  53706-1390}
\author{E.~Bertin}
\affiliation{CNRS, UMR 7095, Institut d'Astrophysique de Paris, F-75014, Paris, France}
\affiliation{Sorbonne Universit\'es, UPMC Univ Paris 06, UMR 7095, Institut d'Astrophysique de Paris, F-75014, Paris, France}
\author{S.~Bhargava}
\affiliation{Department of Physics and Astronomy, Pevensey Building, University of Sussex, Brighton, BN1 9QH, UK}
\author{D.~Brooks}
\affiliation{Department of Physics \& Astronomy, University College London, Gower Street, London, WC1E 6BT, UK}
\author{E.~Buckley-Geer}
\affiliation{Fermi National Accelerator Laboratory, P. O. Box 500, Batavia, IL 60510, USA}
\author{D.~L.~Burke}
\affiliation{Kavli Institute for Particle Astrophysics \& Cosmology, P. O. Box 2450, Stanford University, Stanford, CA 94305, USA}
\affiliation{SLAC National Accelerator Laboratory, Menlo Park, CA 94025, USA}
\author{M.~Carrasco~Kind}
\affiliation{Department of Astronomy, University of Illinois at Urbana-Champaign, 1002 W. Green Street, Urbana, IL 61801, USA}
\affiliation{National Center for Supercomputing Applications, 1205 West Clark St., Urbana, IL 61801, USA}
\author{J.~Carretero}
\affiliation{Institut de F\'{\i}sica d'Altes Energies (IFAE), The Barcelona Institute of Science and Technology, Campus UAB, 08193 Bellaterra (Barcelona) Spain}
\author{F.~J.~Castander}
\affiliation{Institut d'Estudis Espacials de Catalunya (IEEC), 08034 Barcelona, Spain}
\affiliation{Institute of Space Sciences (ICE, CSIC),  Campus UAB, Carrer de Can Magrans, s/n,  08193 Barcelona, Spain}
\author{C.~Chang}
\affiliation{Department of Astronomy and Astrophysics, University of Chicago, Chicago, IL 60637, USA}
\affiliation{Kavli Institute for Cosmological Physics, University of Chicago, Chicago, IL 60637, USA}
\author{M.~Costanzi}
\affiliation{INAF-Osservatorio Astronomico di Trieste, via G. B. Tiepolo 11, I-34143 Trieste, Italy}
\affiliation{Institute for Fundamental Physics of the Universe, Via Beirut 2, 34014 Trieste, Italy}
\author{L.~N.~da Costa}
\affiliation{Laborat\'orio Interinstitucional de e-Astronomia - LIneA, Rua Gal. Jos\'e Cristino 77, Rio de Janeiro, RJ - 20921-400, Brazil}
\affiliation{Observat\'orio Nacional, Rua Gal. Jos\'e Cristino 77, Rio de Janeiro, RJ - 20921-400, Brazil}
\author{T.~M.~Davis}
\affiliation{School of Mathematics and Physics, University of Queensland,  Brisbane, QLD 4072, Australia}
\author{S.~Desai}
\affiliation{Department of Physics, IIT Hyderabad, Kandi, Telangana 502285, India}
\author{H.~T.~Diehl}
\affiliation{Fermi National Accelerator Laboratory, P. O. Box 500, Batavia, IL 60510, USA}
\author{P.~Doel}
\affiliation{Department of Physics \& Astronomy, University College London, Gower Street, London, WC1E 6BT, UK}
\author{A.~Drlica-Wagner}
\affiliation{Fermi National Accelerator Laboratory, P. O. Box 500, Batavia, IL 60510, USA}
\affiliation{Kavli Institute for Cosmological Physics, University of Chicago, Chicago, IL 60637, USA}
\affiliation{Department of Astronomy and Astrophysics, University of Chicago, Chicago, IL 60637, USA}
\author{J.~Estrada}
\affiliation{Fermi National Accelerator Laboratory, P. O. Box 500, Batavia, IL 60510, USA}
\author{S.~Everett}
\affiliation{Santa Cruz Institute for Particle Physics, Santa Cruz, CA 95064, USA}
\author{A.~E.~Evrard}
\affiliation{Department of Astronomy, University of Michigan, Ann Arbor, MI 48109, USA}
\affiliation{Department of Physics, University of Michigan, Ann Arbor, MI 48109, USA}
\author{E.~Fernandez}
\affiliation{Institut de F\'{\i}sica d'Altes Energies (IFAE), The Barcelona Institute of Science and Technology, Campus UAB, 08193 Bellaterra (Barcelona) Spain}
\author{D.~A.~Finley}
\affiliation{Fermi National Accelerator Laboratory, P. O. Box 500, Batavia, IL 60510, USA}
\author{B.~Flaugher}
\affiliation{Fermi National Accelerator Laboratory, P. O. Box 500, Batavia, IL 60510, USA}
\author{P.~Fosalba}
\affiliation{Institut d'Estudis Espacials de Catalunya (IEEC), 08034 Barcelona, Spain}
\affiliation{Institute of Space Sciences (ICE, CSIC),  Campus UAB, Carrer de Can Magrans, s/n,  08193 Barcelona, Spain}
\author{J.~Frieman}
\affiliation{Fermi National Accelerator Laboratory, P. O. Box 500, Batavia, IL 60510, USA}
\affiliation{Kavli Institute for Cosmological Physics, University of Chicago, Chicago, IL 60637, USA}
\author{E.~Gaztanaga}
\affiliation{Institut d'Estudis Espacials de Catalunya (IEEC), 08034 Barcelona, Spain}
\affiliation{Institute of Space Sciences (ICE, CSIC),  Campus UAB, Carrer de Can Magrans, s/n,  08193 Barcelona, Spain}
\author{D.~W.~Gerdes}
\affiliation{Department of Astronomy, University of Michigan, Ann Arbor, MI 48109, USA}
\affiliation{Department of Physics, University of Michigan, Ann Arbor, MI 48109, USA}
\author{D.~Gruen}
\affiliation{Department of Physics, Stanford University, 382 Via Pueblo Mall, Stanford, CA 94305, USA}
\affiliation{Kavli Institute for Particle Astrophysics \& Cosmology, P. O. Box 2450, Stanford University, Stanford, CA 94305, USA}
\affiliation{SLAC National Accelerator Laboratory, Menlo Park, CA 94025, USA}
\author{R.~A.~Gruendl}
\affiliation{Department of Astronomy, University of Illinois at Urbana-Champaign, 1002 W. Green Street, Urbana, IL 61801, USA}
\affiliation{National Center for Supercomputing Applications, 1205 West Clark St., Urbana, IL 61801, USA}
\author{G.~Gutierrez}
\affiliation{Fermi National Accelerator Laboratory, P. O. Box 500, Batavia, IL 60510, USA}
\author{S.~R.~Hinton}
\affiliation{School of Mathematics and Physics, University of Queensland,  Brisbane, QLD 4072, Australia}
\author{D.~L.~Hollowood}
\affiliation{Santa Cruz Institute for Particle Physics, Santa Cruz, CA 95064, USA}
\author{K.~Honscheid}
\affiliation{Center for Cosmology and Astro-Particle Physics, The Ohio State University, Columbus, OH 43210, USA}
\affiliation{Department of Physics, The Ohio State University, Columbus, OH 43210, USA}
\author{D.~J.~James}
\affiliation{Center for Astrophysics $\vert$ Harvard \& Smithsonian, 60 Garden Street, Cambridge, MA 02138, USA}
\author{S.~Kent}
\affiliation{Fermi National Accelerator Laboratory, P. O. Box 500, Batavia, IL 60510, USA}
\affiliation{Kavli Institute for Cosmological Physics, University of Chicago, Chicago, IL 60637, USA}
\author{E.~Krause}
\affiliation{Department of Astronomy/Steward Observatory, University of Arizona, 933 North Cherry Avenue, Tucson, AZ 85721-0065, USA}
\author{K.~Kuehn}
\affiliation{Australian Astronomical Optics, Macquarie University, North Ryde, NSW 2113, Australia}
\affiliation{Lowell Observatory, 1400 Mars Hill Rd, Flagstaff, AZ 86001, USA}
\author{H.~Lin}
\affiliation{Fermi National Accelerator Laboratory, P. O. Box 500, Batavia, IL 60510, USA}
\author{M.~A.~G.~Maia}
\affiliation{Laborat\'orio Interinstitucional de e-Astronomia - LIneA, Rua Gal. Jos\'e Cristino 77, Rio de Janeiro, RJ - 20921-400, Brazil}
\affiliation{Observat\'orio Nacional, Rua Gal. Jos\'e Cristino 77, Rio de Janeiro, RJ - 20921-400, Brazil}
\author{M.~March}
\affiliation{Department of Physics and Astronomy, University of Pennsylvania, Philadelphia, PA 19104, USA}
\author{J.~L.~Marshall}
\affiliation{George P. and Cynthia Woods Mitchell Institute for Fundamental Physics and Astronomy, and Department of Physics and Astronomy, Texas A\&M University, College Station, TX 77843,  USA}
\author{P.~Melchior}
\affiliation{Department of Astrophysical Sciences, Princeton University, Peyton Hall, Princeton, NJ 08544, USA}
\author{F.~Menanteau}
\affiliation{Department of Astronomy, University of Illinois at Urbana-Champaign, 1002 W. Green Street, Urbana, IL 61801, USA}
\affiliation{National Center for Supercomputing Applications, 1205 West Clark St., Urbana, IL 61801, USA}
\author{R.~Miquel}
\affiliation{Instituci\'o Catalana de Recerca i Estudis Avan\c{c}ats, E-08010 Barcelona, Spain}
\affiliation{Institut de F\'{\i}sica d'Altes Energies (IFAE), The Barcelona Institute of Science and Technology, Campus UAB, 08193 Bellaterra (Barcelona) Spain}
\author{R.~L.~C.~Ogando}
\affiliation{Laborat\'orio Interinstitucional de e-Astronomia - LIneA, Rua Gal. Jos\'e Cristino 77, Rio de Janeiro, RJ - 20921-400, Brazil}
\affiliation{Observat\'orio Nacional, Rua Gal. Jos\'e Cristino 77, Rio de Janeiro, RJ - 20921-400, Brazil}
\author{F.~Paz-Chinch\'{o}n}
\affiliation{Institute of Astronomy, University of Cambridge, Madingley Road, Cambridge CB3 0HA, UK}
\affiliation{National Center for Supercomputing Applications, 1205 West Clark St., Urbana, IL 61801, USA}
\author{A.~A.~Plazas}
\affiliation{Department of Astrophysical Sciences, Princeton University, Peyton Hall, Princeton, NJ 08544, USA}
\author{A.~Roodman}
\affiliation{Kavli Institute for Particle Astrophysics \& Cosmology, P. O. Box 2450, Stanford University, Stanford, CA 94305, USA}
\affiliation{SLAC National Accelerator Laboratory, Menlo Park, CA 94025, USA}
\author{M.~Sako}
\affiliation{Department of Physics and Astronomy, University of Pennsylvania, Philadelphia, PA 19104, USA}
\author{E.~Sanchez}
\affiliation{Centro de Investigaciones Energ\'eticas, Medioambientales y Tecnol\'ogicas (CIEMAT), Madrid, Spain}
\author{V.~Scarpine}
\affiliation{Fermi National Accelerator Laboratory, P. O. Box 500, Batavia, IL 60510, USA}
\author{M.~Schubnell}
\affiliation{Department of Physics, University of Michigan, Ann Arbor, MI 48109, USA}
\author{S.~Serrano}
\affiliation{Institut d'Estudis Espacials de Catalunya (IEEC), 08034 Barcelona, Spain}
\affiliation{Institute of Space Sciences (ICE, CSIC),  Campus UAB, Carrer de Can Magrans, s/n,  08193 Barcelona, Spain}
\author{I.~Sevilla-Noarbe}
\affiliation{Centro de Investigaciones Energ\'eticas, Medioambientales y Tecnol\'ogicas (CIEMAT), Madrid, Spain}
\author{J.~Allyn.~Smith}
\affiliation{Austin Peay State University, Dept. Physics, Engineering and Astronomy, P.O. Box 4608 Clarksville, TN 37044, USA}
\author{M.~Smith}
\affiliation{School of Physics and Astronomy, University of Southampton,  Southampton, SO17 1BJ, UK}
\author{E.~Suchyta}
\affiliation{Computer Science and Mathematics Division, Oak Ridge National Laboratory, Oak Ridge, TN 37831}
\author{G.~Tarle}
\affiliation{Department of Physics, University of Michigan, Ann Arbor, MI 48109, USA}
\author{M.~A.~Troxel}
\affiliation{Department of Physics, Duke University Durham, NC 27708, USA}
\author{D.~L.~Tucker}
\affiliation{Fermi National Accelerator Laboratory, P. O. Box 500, Batavia, IL 60510, USA}
\author{A.~R.~Walker}
\affiliation{Cerro Tololo Inter-American Observatory, NSF's National Optical-Infrared Astronomy Research Laboratory, Casilla 603, La Serena, Chile}
\author{W.~Wester}
\affiliation{Fermi National Accelerator Laboratory, P. O. Box 500, Batavia, IL 60510, USA}
\author{R.D.~Wilkinson}
\affiliation{Department of Physics and Astronomy, Pevensey Building, University of Sussex, Brighton, BN1 9QH, UK}
\author{J.~Zuntz}
\affiliation{Institute for Astronomy, University of Edinburgh, Edinburgh EH9 3HJ, UK}

\collaboration{1000}{(DES Collaboration)}


\begin{abstract}
    We present a measurement of the Hubble constant $H_0$ using the gravitational wave (GW) event GW190814, which resulted from the coalescence of a 23 $M_\odot$ black hole with a 2.6 $M_\odot$ compact object, as a standard siren.
    No compelling electromagnetic counterpart with associated host galaxy has been identified for this event, thus our analysis accounts for thousands of potential host galaxies within a statistical framework. The redshift information is obtained from the photometric redshift (photo-$z$) catalog from the Dark Energy Survey. The luminosity distance is provided by the gravitational wave sky map published by the LIGO/Virgo Collaboration. Since this GW event has the second--smallest sky localization area after GW170817, GW190814 is likely to provide the best constraint on cosmology from a single standard siren without identifying an electromagnetic counterpart. Our analysis uses photo-$z$ probability distribution functions and corrects for photo-$z$ biases. We also reanalyze the binary--black hole GW170814 within this updated framework. We explore how our findings impact the $H_0$ constraints from GW170817, the only GW merger associated with a unique host galaxy, and therefore the most powerful standard siren to date. From a combination of GW190814, GW170814 and GW170817, our analysis yields {$H_0 =  72.0^{+ 12}_{- 8.2 }~{\rm km~s^{-1}~Mpc^{-1}}$ }(68\% Highest Density Interval, HDI) for a prior in $H_0$ uniform between $[20,140]~{\rm km~s^{-1}~Mpc^{-1}}$. The addition of GW190814 and GW170814 to GW170817 improves the 68\% HDI from GW170817 alone by {$\sim 18\%$}, showing how well--localized mergers without counterparts can provide a {significant} contribution to standard siren measurements, provided that a complete galaxy catalog is available at the location of the event.
\end{abstract}

\keywords{catalogs --- cosmology: observations --- gravitational waves --- surveys}

\reportnum{DES-2020-0548}
\reportnum{FERMILAB-PUB-20-216-AE}

\section{Introduction}

The first detection of gravitational waves (GW) from a binary--black--hole merger (GW150914; \citealt{GW150914}), and only two years later the first detection of a binary--neutron--star merger (GW170817; \citealt{ligobns}) with associated electromagnetic counterpart (\citealt{MMApaper,marcelle17,arcavi,Coulter1556,lipunov,tanvir,valenti}), have generated tremendous excitement amongst the astrophysics community. One of the most appealing applications of gravitational wave detections is in cosmological analyses. The GW signal from mergers of neutron--star and black--hole binary systems are in fact absolute distance indicators, and can be used as ``standard sirens'', as first proposed in \citet{schutz}. The luminosity distance to the source can be inferred from the gravitational wave signal, and if a redshift measurement is also available (for example, through identification of the host galaxy), we can measure the present rate of expansion of the Universe $H_0$ via the distance--redshift relation. 

New and independent measurements of the Hubble constant are of great interest to cosmology. Measurements obtained with type Ia Supernovae (SN Ia) and inferred from the cosmic microwave background (CMB) (e.g. \citealt{Riess_2019,planck18,Freedman_2019}) have each reached the $1-2\%$ precision level, although the latest measurements {from SN Ia \citep{Riess_2019} and CMB \citep{planck18} disagree at the 4.4$\sigma$ level.} Independent, precise and accurate measurements of $H_0$ could help clarify whether the tension arises from beyond--$\Lambda$CDM physics or unknown systematics (\citealt{2017NatAs...1E.169F,2018JCAP...09..025M,verde}).

The standard siren methodology was applied for the first time to the binary--neutron--star (BNS) merger GW170817 \citep{2017Natur.551...85A}, thanks to the association of its electromagnetic (EM) counterpart to the nearby host galaxy NGC 4993 (e.g. \citealt{palmese}). However, no other compelling counterparts to GW events have been identified to date (e.g. \citealt{growth,vieira,morgan2020constraints,Garcia:2020smy}). 
Events without counterparts can also be used for cosmological analyses, using a statistical approach first proposed in \citet{schutz}. If a complete catalog of potential host galaxies exists within the event localization region, their redshift distribution can provide the redshift information needed to infer cosmological parameters from the distance--redshift relation. We refer to this method as the ``dark'' or ``statistical'' standard siren method, as opposed to the ``bright'' or ``counterpart'' case. \citet{delpozzo} and \citet{chen17} provide a Bayesian framework that enables a measurement of $H_0$ with the statistical approach. This framework is implemented in \citet{fishbach} using GW170817, and in \cite*{darksiren1}, where we measure the Hubble constant using a binary--black--hole (GW170814; \citealt{gw170814}) for the first time. \citet{LVC_O2_StS} combine dark sirens from all of the binary black holes detected during O1 and O2, but find that the method could only bring a $\sim 7\%$ improvement to the GW170817 bright siren constraint due to the lack of complete public galaxy catalogs in the regions of interest {and to the poor localization of most of the events}. Eventually, a large sample of events, combined with wide field on-going and upcoming galaxy surveys (e.g. LSST, DESI, 4MOST), will enable precise cosmological measurements with dark and bright standard sirens (e.g. \citealt{2005ApJ...629...15H,macleod,2010ApJ...725..496N,delpozzo,2013arXiv1307.2638N,2017PhRvD..96j1303N,chen17,2018PhRvL.121b1303V,nair,2019PhRvD.100j3523M,2019PhRvL.122f1105F,palmese_WP,yu2020hunting}). In particular, \citet{chen17} forecast a 2\% precision within 5 years for bright standard sirens detected by LIGO/Virgo. \citet{nair} predict a $\sim 7\%$ measurement with 25 binary black hole (BBH) events from {next generation GW detectors}, while \citet{yu2020hunting} anticipate that a $1-4\%$ measurement {will be} possible with LIGO/Virgo BBHs if these mergers originate in groups and clusters of galaxies. Interesting cosmological constraints can also be placed using gravitational wave compact binary mergers and large galaxy surveys by measuring their peculiar velocity power spectrum, along with their overdensity and cross--correlation power spectra, as proposed in \citet{palmese20}.

With the promises of gravitational wave cosmology in mind, the Dark Energy Survey (DES) collaboration and external collaborators launched the DES gravitational waves (DESGW) program. As part of this program, we seek optical emission from LIGO/Virgo events using the Dark Energy Camera (DECam; \citealt{flaugher}). Our searches resulted in the independent discovery of the kilonova associated with GW170817 \citep{marcelle17}, and in some of the most stringent limits for optical emission from binary--black--hole (\citealt{2016ApJ...823L..33S,2016ApJ...826L..29C,doctor}) and other compact object \citep{morgan2020constraints} mergers.

In this article we measure $H_0$ using the gravitational wave event GW190814 (\citealt{gcn_skymap,190814_paper}), which resulted from the inspiral and merger of 
a 23 $M_\odot$ black hole with a 2.6 $M_\odot$ compact object at $241^{+41}_{-45}$ Mpc (90\% credible interval; \citealt{190814_paper}). 
The secondary component was either the lightest black hole or the heaviest neutron star ever found in a binary system. This event is particularly interesting for a dark standard siren analysis because counterpart searches have not identified a convincing counterpart thus far (e.g. \citealt{morgan2020constraints,growth,vieira,Watson20,Gomez20,Ackley20}), and because its localization volume is the second--smallest after GW170817. Additionally, the event localization region falls within the DES footprint, making DES galaxy catalogs an ideal sample for a measurement of $H_0$. %
{\citet{190814_paper}, which became publicly available almost concurrently with this work, provide a brief description of a dark siren analysis for this event using an inhomogeneous galaxy sample, mostly incomplete at the redshift range of interest. In this work, we provide a detailed standard siren analysis for GW190814 that takes advantage of a more complete galaxy catalog.} Compared to other previous works on dark standard sirens, we also improve upon the treatment of the photometric redshifts (photo-$z$'s) by taking into account their full probability distribution function (PDF) rather than a Gaussian approximation of the former, and by correcting for photo-$z$ biases in the data. This methodology ensures a more accurate recovery of the true redshift distribution of galaxies, as it has been extensively studied in the literature, particularly within the context of weak gravitational lensing analyses (e.g. \citealt{lima,oyaizu,cunha}). We also reanalyze GW170814 using DES galaxies within this updated framework and provide a combined $H_0$ measurement using three events.

{We describe our dataset in \S\ref{data} and the methods used in \S\ref{method}. Our results and discussion follow in \S\ref{results}, and the conclusions are in \S\ref{conclusions}. We assume a flat $\Lambda$CDM cosmology with $\Omega_m =0.3$ and $H_0$ values in the $20-140~{\rm km~s^{-1}~Mpc^{-1}}$ range. When not otherwise stated, quoted error bars represent the 68\%\ credible interval (CI).}

\section{Data}\label{data}

\begin{figure*}
\centering
\begin{minipage}[c]{0.45\linewidth}
\includegraphics[width=1\linewidth]{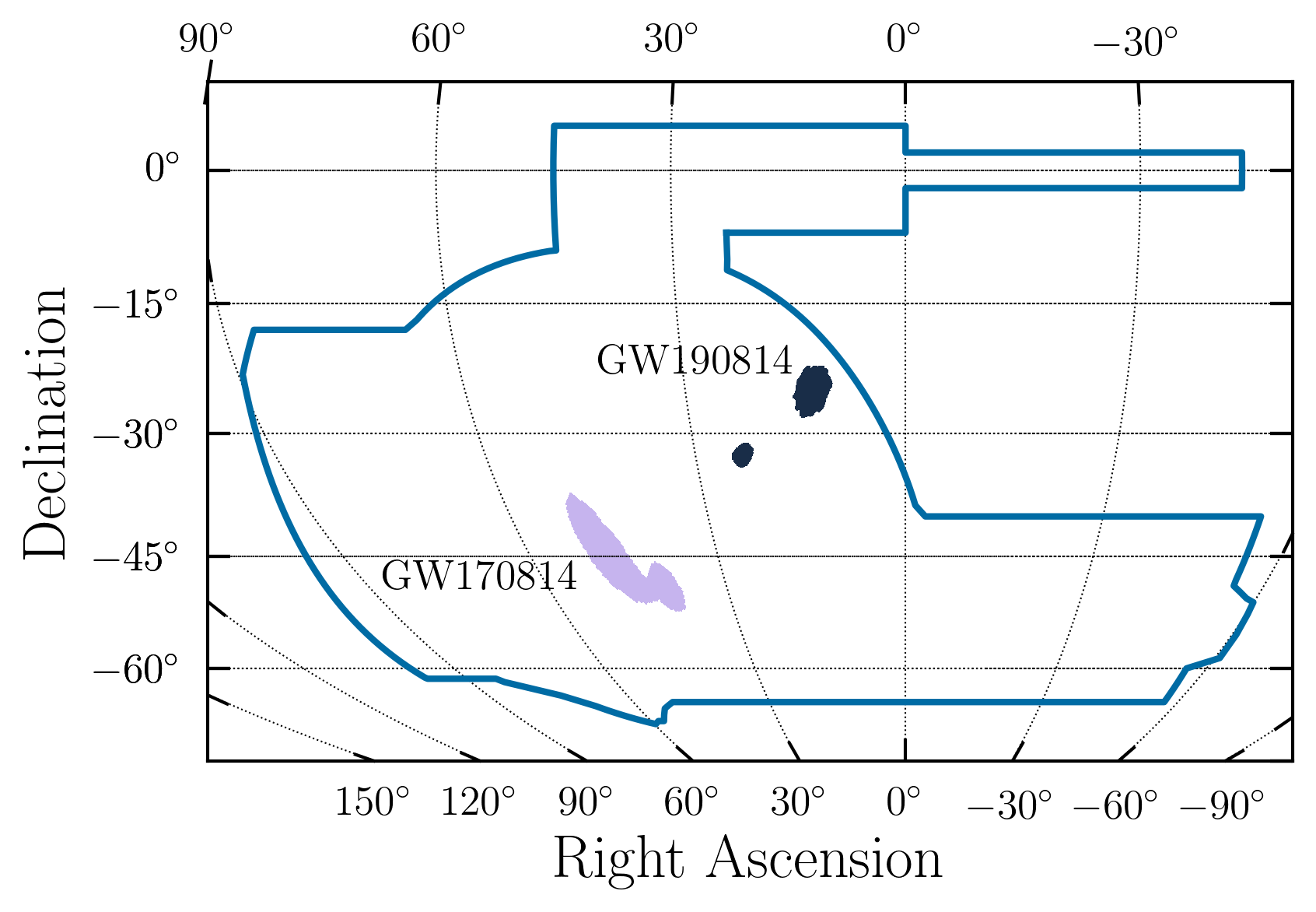}
\end{minipage}{}
\begin{minipage}[c]{0.45\linewidth}
\includegraphics[width=0.9\linewidth,trim=0 0 0.1 0.3, clip]{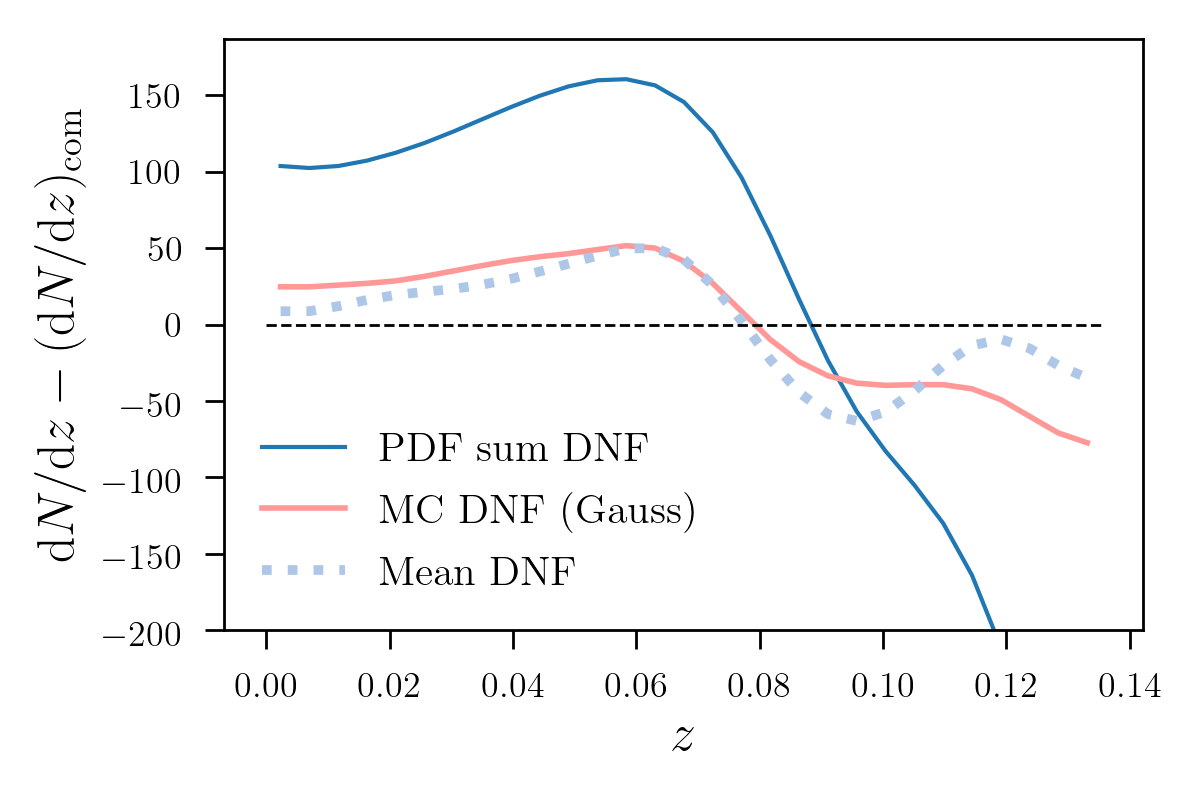}\\
\includegraphics[width=0.9\linewidth,trim=0 0 0.1 0.3, clip]{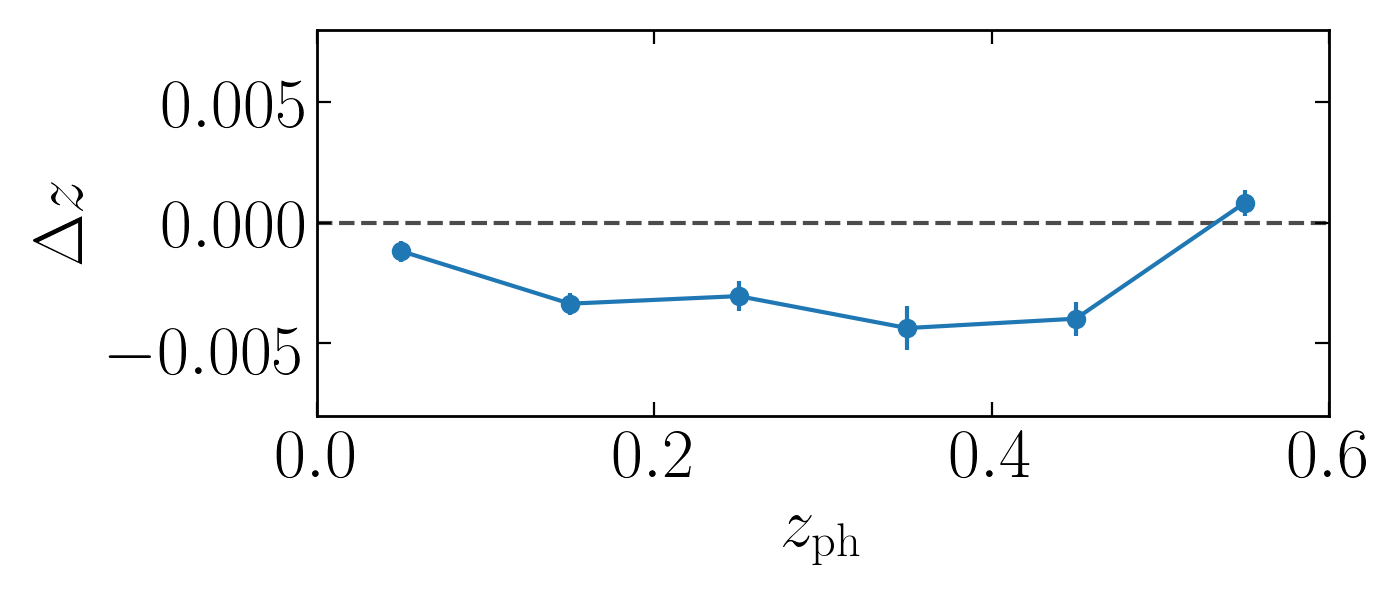}
\end{minipage}{}

\caption{\emph{Left:} LIGO/Virgo GW dark standard sirens analyzed in this paper, where the shaded regions represent the 90\% CI localization from the sky maps. {For GW190814, the localization comes from the \texttt{LALInference} sky map.} The blue contour represents the DES footprint.
\emph{Top right:} {Difference between several inferred redshift distributions $\de N/\de z$ and the distribution $(\de N/\de z)_{\rm com}$ for a uniform-number-density case with $H_0=70~{\rm km~s^{-1}~Mpc^{-1}}$. {This difference is shown to highlight the presence of overdensities and underdensities in the $\de N/\de z$ distribution.} The redshifts have been inferred with the DNF method using DES photometry.} The redshift range shown is the one relevant for an $H_0$ prior extending out to $140~{\rm km~s^{-1}~Mpc^{-1}}$. {The dotted line represents the distribution obtained when considering the mean photo--$z$ value of each galaxy as provided by DNF}. The pink curve is given by Monte Carlo (MC) sampling the photo-$z$ PDF for each galaxy, assuming that this is a Gaussian centered on the mean value (from the dotted line) with {standard deviation} given by the 1$\sigma$ uncertainty provided by DNF. 
The blue solid curve is a sum of the full, non-Gaussian, redshift PDFs of each galaxy, and is used in our final results. \emph{Bottom right:} DNF photo-$z$ bias $\Delta z = \langle z_{\rm sp} \rangle - \langle z_{\rm ph} \rangle$ in bins of photo-$z$ measured on a validation sample of $\sim 140,000$ galaxies with spectroscopic data in the DES footprint. The photo-$z$ estimator used to compute the bias is the mean DNF PDF value.}
\label{fig:data}
\end{figure*}

\subsection{The LIGO/Virgo GW data}\label{sec:GWdata}

The gravitational wave data used in this work come from the publicly available sky maps. First we use the \texttt{LALInference} sky map \citep{gcn_skymap}\footnote{\url{https://gracedb.ligo.org/superevents/S190814bv/view/}} released in August 2019. The area of the sky enclosing 90\% of the localization probability is 23 deg$^2$, and it is shown in Figure \ref{fig:data} together with the GW170814 90\% CI contours and the DES footprint (in blue). Angular and distance probabilities are provided in HEALPix \citep{healpix} pixels, where the distance probability distribution is approximated by a Gaussian along each line of sight. The maximum probability pixel is centered at RA, Dec = $(12.832,-25.241)$ deg. Marginalized over the whole sky, the luminosity distance has a mean value of 267 Mpc and a standard deviation of 52 Mpc in the Gaussian approximation. The resulting 90$\%$ CI comoving volume is $9.2\times 10^4$ Mpc$^3$, which is two orders of magnitude larger than the GW170817 volume, but two orders of magnitude smaller than GW170814.
{We derive the redshift range of interest for this analysis by considering the highest value of $H_0$ in our prior range (i.e. 140 $ ~{\rm km~s^{-1}~Mpc^{-1}}$) with a flat $\Lambda$CDM cosmology having $\Omega_m=0.3$, and the high limit of the luminosity distance at 90 and 99.7\% CI.  These two CI limits correspond to $z=0.15$ and $0.18$, implying that our analysis is sensitive to galaxies at redshifts below these values.}

{After submission of this manuscript, we added results using the sky map released by the LVC in July 2020\footnote{\url{https://dcc.ligo.org/LIGO-P2000230/public}} {(hereafter called the higher modes map)}. This map provides more precise parameter estimation thanks to the higher order modes detected in the offline analysis by the LVC \citep{190814_paper}. As a result, the 90\% CI localization area becomes 18.5 deg$^2$, and the luminosity distance is $241\pm 26$ Mpc (mean and standard deviation). The updated 90$\%$ CI comoving volume is about a third of the one from the \texttt{LALInference} map, $3.2\times 10^4$ Mpc$^3$.}

\subsection{The DES data}\label{sec:des}

The DES\footnote{\url{www.darkenergysurvey.org}}(\citealt{descollaboration,2016MNRAS.460.1270D}) is an optical--near-infrared survey 
that imaged 5000 ${\rm deg}^2$ of the South Galactic Cap over six years in the $grizY$ bands. DES used the $\sim3$ $\textrm{deg}^2$ DECam, mounted on the Blanco 4-m telescope at the Cerro Tololo Inter-American Observatory (CTIO) in Chile. In our analysis we use data from the first 3 years of observations (``Y3''; \citealt{dr1}; September 2013 -- February 2016).
 
{Our reduced data products are from the DES Data Management (DESDM) pipeline (\citealt{2018PASP..130g4501M}), which includes calibration of the single-epoch images, background subtraction, co--addition of the background-subtracted images, and cutting the co--added images into tiles. The source catalogue was created with \textsc{Source Extractor (SExtractor}, \citealt{sextractor}), which detects objects on the $riz$ co-added images.} For galaxies the median $10\sigma$ limiting magnitudes in the Y3 data are $g = 24.33$, $r = 24.08$, $i = 23.44$, $z = 22.69$, and $Y = 21.44$ mag (\citealt{dr1}). The data were carefully selected to produce a Y3 ``gold'' catalog, as described in Sevilla-Noarbe et al., in prep.
In this work we use the gold catalog, and the photometry is derived through the Single--Object Fitting (SOF) pipeline that relies on the \texttt{ngmix} code.\footnote{\url{https://github.com/esheldon/ngmix}}
The SOF fluxes have been used to compute photometric redshifts using the Directional Neighborhood Fitting (DNF; \citealt{dnf}) method, as described in more detail in the next subsection.

Galaxy properties for the Y3 sample (Palmese et al., in prep.) are derived using the SOF photometry, and by fixing the redshift to the DNF photo-$z$ mean value, or spectroscopic redshift where available. In particular, stellar mass and absolute magnitudes are derived  through a broadband Spectral Energy Distribution (SED) fitting of galaxy magnitudes with \textsc{LePhare} (\citealt{arnouts}, \citealt{ilbertlephare}). Estimates of the galaxy properties used here from DES data alone have been tested and studied in several DES works (\citealt{palmese16,Etherington,palmese18}). We add a 0.05 systematic uncertainty in quadrature to the magnitudes, to account for systematic uncertainties in magnitude estimation and model variance. {The templates used for the galaxies' SED fitting are }the simple stellar populations (SSP) from \citet{bc03}, with three metallicities ($0.2~Z_\odot$, $Z_\odot$ and $2.5~Z_\odot$), a \citet{chabrier} Initial Mass Function (IMF) and a Milky Way \citep{allen} extinction law with five different values between 0 and 0.5 for the $E(B-V)$ reddening. The star formation history (SFH) chosen is exponentially declining with the age of the galaxy $t$ as ${\rm e}^{-t/\tau}$ with $\tau=0.1,0.3,1,2,3,5,10,15$ and $30$ Gyr. 

{As discussed in \citealt{dr1} the Y3 gold catalog source list 
is $95\%$ complete for galaxies within our apparent magnitude limit of $r<23.35$.}
{By converting the apparent magnitude source completeness into a completeness in redshift intervals using the DNF photo--$z$'s}, in \cite*{darksiren1} we found that our sample with $r<23.35$ is $>93\%$ complete within $z<0.26$. As a result, our catalog is highly complete over the redshift range of interest, therefore we do not apply completeness corrections. However, the sample is still magnitude--limited, implying that intrinsically fainter galaxies will be more easily observed at the lowest redshifts. We overcome this problem by defining a volume-limited sample, obtained by applying a luminosity cut.
Following \cite{Pozzetti10} and \cite{Hartley13}, we identify galaxies that are bright enough to be complete and representative of the real galaxy population. In order to facilitate a comparison with GW170814, we adopt the same luminosity cut used in \cite*{darksiren1} because such a limit is valid in the range $0<z<0.26$, which includes the redshifts of interest here. This cut is made at $-17.2$ {(computed assuming $H_0=70~{\rm km~s^{-1}~Mpc^{-1}}$, and it is rescaled accordingly for other $H_0$ values)} in $r$-band absolute magnitude and $\sim 3.8 \times 10^8~{\rm M}_\odot$ in stellar mass. As in  \cite*{darksiren1}, the sample remaining after the cut contains $> 77\%$ of the stellar mass in
the volume, assuming the galaxies follow a Schechter stellar mass function with the best fit values from \citet{weigel}.

{In the next subsection and in particular for Figure \ref{fig:pdfs},  we have used the \texttt{LALInference} map.
As the higher modes map has a smaller spatial localization and smaller luminosity distance uncertainties, 
the $\sim 1,800$ galaxy sample encompassing the 90\% CI volume using it is a strict subset of the $\sim 3,800$ galaxy sample selected using the \texttt{LALInference}.}

\subsubsection{The redshift catalog}

The DNF method applied to Y3 data provides redshift information for each galaxy in the form of a PDF, from which we compute a mean redshift, and half of the central 68th percentile width. {Our method requires having individual galaxies' PDFs as will be clear in Section \ref{method}. We use the full PDF for each galaxy, and contrast the results with a Gaussian approximation of the PDF using mean redshifts with the $1\sigma$ values.} Where available, we use public spectroscopic redshifts to complement the sample.

The final galaxy redshift distributions are shown in {the top right panel of} Figure \ref{fig:data}. There are 2,684 galaxies in the GW190814 90\% probability region supporting $H_0$ values out to $140 ~{\rm km~s^{-1}~Mpc^{-1}}$ ($z \sim 0.135$). Of these, $\sim 700$ have available spectroscopic redshifts from 2dF, 2dFLens and 6dF \citep{2001MNRAS.328.1039C,2009MNRAS.399..683J}. The top right plot in Figure \ref{fig:data} shows the difference between several inferred redshift distributions $\de N/\de z$ and {the distribution $(\de N/\de z)_{\rm com}$} for a uniform--number--density case (i.e. a distribution which is uniform in \emph{comoving} volume) with $H_0=70~{\rm km~s^{-1}~Mpc^{-1}}$. The different lines show the results from different redshift estimators. {The dotted line represents the $\de N/\de z$ given by the mean photo--$z$ value of each galaxy as provided by DNF. The pink curve is given by Monte Carlo (MC) sampling a Gaussian approximation of the photo-$z$ PDF for each galaxy. In this case, we assume that the PDF is a Gaussian with width given by the 1$\sigma$ uncertainty provided by DNF.} 
The blue solid curve is a sum of the full, non-Gaussian, redshift PDFs of each galaxy.

We have tested the aforementioned redshift estimators using a spectroscopic sample of $140,000$ galaxies matched to DES Y3 objects as a validation sample \citep{julia}. This sample does not contain any galaxies used for training DNF. The metric that we choose to evaluate the performance of the photo--$z$ algorithm in recovering the redshift distribution 
is:
\begin{equation}
\bar{\Delta N(z)} (\%)=    \sum_k \frac{|N_{ph}(z_k)-N_{sp}(z_k)|}{N_{sp}(z_k)}\times 100 \, ,
\end{equation}
where the sum extends over the redshift bins of interest, and $N_{ph}(z_k)$ and $N_{sp}(z_k)$ are the photometric and spectroscopic sample galaxy counts in the $k$-th bin, respectively. {In the case of the full photo--$z$ PDF, $N_{ph}(z_k)$ is computed by summing the contributions of single galaxies' PDFs in each bin $k$. We consider redshift bins of width 0.05 out to $z<0.6$.} Compared to using the mean photo-$z$ values for $N_{ph}(z_k)$, this metric is improved by more than a factor of 2 when $N_{ph}(z_k)$ is summed by integrating the PDF of all galaxies over the $k$'th redshift bin. Using MICE simulations \citep{Fosalba_2015,Crocce_2015}, we also find that the $\bar{\Delta N(z)}$ metric is significantly reduced (again by a factor of $\sim 2$) when using the full PDF versus the mean DNF values. The magnitude of the improvement is equivalent to adding $u$ band to the measurements. This is due to the fact that the full PDF captures the effect of color-redshift degeneracies. {On the other hand, these degeneracies are not well captured when only a point estimate such as the mean is taken into account, or when the PDF is approximated by a Gaussian} (\citealt{Buchs2019}; Gschwend et al., in prep.). We conclude that accounting for full PDF is more accurate than using point estimates, as they are able to best reproduce the redshift distribution of galaxies. This conclusion is broadly supported by several works on photometric surveys (e.g. \citealt{lima,oyaizu,cunha}).

Using the spectroscopic validation sample, we estimate the extent of systematic biases that propagate to a bias in the $H_0$ posterior. We find that in the redshift range of interest for this work, $z<0.3$, the bias $\Delta z = \langle z_{\rm sp} \rangle - \langle z_{\rm ph} \rangle$ in bins of photo--$z$ {with width 0.1} is of the order $\sim 0.002$. The photo-$z$ estimator used to compute this bias is the mean DNF value. We marginalize over the redshift dependent bias shown in the bottom right panel of Figure \ref{fig:data}.

We notice that the redshift distributions show the large ``galaxy wall'' already seen for GW170814 (see Fig.1 of \citealt*{darksiren1}), spanning most of the DES footprint and confirmed by spectroscopic data, at $z\sim 0.06$. This overdensity is also present in Bayesian Photometric Redshift (BPZ; \citealt{benitez}) redshifts, which is a template--based method, proving that the structure is not a result of machine learning training in DNF. 




\section{Method}\label{method}

In this work, we follow the approach of \citet{chen17}, and slightly adapt it for our purposes, similarly to \citet*{darksiren1}.
The posterior probability of $H_0$ given the GW data $d_{\rm GW}$ from a single event detection and EM data $d_{\rm EM}$ from a galaxy survey can be written as:
\begin{equation}
p(H_0|d_{\rm GW},d_{\rm EM})  \propto p(d_{\rm GW},d_{\rm EM}|H_0)p(H_0)\, ,
\label{eq:posterior}
\end{equation}
as follows from Bayes' Theorem. The joint likelihood $p(d_{\rm GW},d_{\rm EM}|H_0)$ can be written as the product of two individual likelihoods, $p(d_{\rm GW}|H_0)$ and $p(d_{\rm EM}|H_0)$, since the GW and EM data are independent. The GW likelihood is marginalized over all variables except for the true luminosity distance $d_L$ and solid angle $\hat{\Omega}_{\rm GW}$ of the GW source, while the EM likelihood explicitly depends on the true redshift $z_i$ and solid angle $\hat{\Omega}_i$ for each galaxy $i$. We refer to $\hat{\Omega}_i$ as ``solid angles'', but these are vectors with the angular position of the source/galaxy as direction, subtending the area ($\sim 3 \times 10^{-3}$ deg$^2$) of each HEALPix pixel. We assume that the GW source is located within one of the galaxies in the galaxy catalog. Therefore, the location and distance of the GW source can be directly related to solid angle $\hat{\Omega}_i$ and the redshift (through the cosmology) of each galaxy. Finally, we marginalize over the choice of galaxy $i$, over the true luminosity distance and over the position of the GW source, and write the joint, marginal likelihood as:

\begin{equation}
\begin{split}
p(d_{\rm GW}, d_{\rm EM}|\{z_j,\hat{\Omega}_j\},H_0) \propto  \sum_i w_i \!\int\!\de d_L \, \de \hat{\Omega}_{\rm GW}\,p(d_{\rm GW}|d_L,\hat{\Omega}_{\rm GW}) \\
\times p(d_{\rm EM}|\{z_j,\hat{\Omega}_j\})\,\delta_D(d_L-d_L(z_i,H_0))
\,\delta_D(\hat{\Omega}_{\rm GW}-\hat{\Omega}_{i})\,,\label{eq:like}
\end{split}
\end{equation}
where {$\{z_j,\hat{\Omega}_j\}$} are all of the galaxies' redshift and solid angle, $w_i$ are weights that represent the relative probability that the $i$th galaxy hosts a GW source, and $\delta_D$ is the Dirac delta function. In principle, the $w_i$ weights would be based on galaxy properties, but as we do not know whether GW host galaxy properties such as star--formation rate or luminosity differ from the general population, here we take the weights to be uniform across all galaxies. {We therefore omit them in the following equations.}

Eq. (\ref{eq:like}) needs to be marginalized over the galaxies' redshifts and sky positions, which require a prior $p(z_i,\Omega_i)$. To first order, galaxies are uniformly distributed in comoving volume $V$, and assuming that our sample is volume--limited within $V_{\rm max}$:
\begin{equation}
\begin{split}
    p(z_i,\hat{\Omega}_i)~ \de z_i~ \de \hat{\Omega}_i & \propto \frac{1}{ V_{\rm max}} \frac{\de^2  V}{\de z_i \de \hat{\Omega}_i} \de z_i~ \de \hat{\Omega}_i \propto \frac{1}{V_{\rm max}} \frac{r^2(z_i)}{H(z_i)} \de z_i ~\de \hat{\Omega}_i\,,
    \end{split}
\end{equation}
where $r$ is the comoving distance to the galaxy. 

If we assume that the galaxies' positions $\{\hat{\Omega}_j\}$ are known with exquisite precision, these can easily be marginalized over with a delta function about the observed values, reducing the marginal EM likelihood to $p(d_{\rm EM}|\{z_j\})$. The assumption on the precision of the galaxies' position is realistic given that spatial probabilities for the GW sources vary significantly over scales that are usually larger than a galaxy's size at the redshifts of interest (in this specific work, they are uniform in each \textsc{HEALPix} pixel). The marginal EM likelihood is then given by:
\begin{equation}
p(d_{\rm EM}|\{z_j\})= \prod_k p(\tilde{z}_{k}|z_k) \, ,\label{eq:emlike}
\end{equation}
{where the EM data $d_{\rm EM}$ here only refers to the measured (photometric or spectroscopic) redshift $\tilde{z}_{k}$ for each galaxy $k$.}
In the photo--$z$ Gaussian approximation case mentioned above, we approximate Eq. (\ref{eq:emlike}) with a product of Gaussian distributions, ${\mathcal N}$, for each galaxy, centred around the observed redshift values $z_{{\rm obs}, k}$ with a width given by the redshift's uncertainty $\sigma_{z,k}$ for each galaxy $k$.

We compute the marginal GW likelihood $p(d_{\rm GW}|d_L,\hat\Omega)$ according to \citet{Singer}:
\begin{equation}
p(d_{\rm GW}|d_L,\hat{\Omega}) \propto p(\hat{\Omega}) \frac{1}{\sqrt{2\pi}\sigma(\hat{\Omega})}{\rm exp}\Big[ - \frac{\big(d_L-\mu (\hat{\Omega}) \big)^2}{2\sigma^2(\hat{\Omega})} \Big] N(\hat{\Omega})\,, \label{eq:dpdvmap} 
\end{equation}
with the sky map providing the position probability $p(\hat{\Omega})$, mean $\mu$, normalization $N$, and scale $\sigma$ at each position.

{We take into account selection effects arising from the detection of GW events {through a normalization factor $\beta(H_0)$, by which we divide the right--hand side of Eq. (\ref{eq:posterior})}. Omitting these effects would bias our inference of $H_0$, in particular because the detection of GW events depends on the Hubble constant, amongst other variables (e.g. \citealt{chen17,mandel,2019PhRvD.100j3523M}).} Following \citet{chen17} and \citet{mandel}, we normalize the likelihood over all possible GW and EM data with a $[\beta(H_0)]^{-1}$ factor. In this work, the galaxy catalog is volume--limited and complete beyond the maximum observable distance for the GW events, so that the normalization factor becomes:
\begin{equation}
\beta(H_0) = \frac{V[d_{L,{\rm GW}}^{\rm max}(H_0)]}{V_{\rm max} (H_0)}\, ,
\end{equation}
where $V[d_{L,{\rm GW}}^{\rm max}(H_0)]$ is the maximum observable volume for the GW events considered. {This volume is computed assuming the distance reach for $30+30~M_\odot$ BBH from \citet{2018LRR....21....3A}, namely 910 and $\sim 1100$ Mpc for O2 and O3, respectively. We have verified that the choice of the masses (and in turn, of the exact value of the distance reach) does not have a significant effect on our final results. In fact, the volume dependence on $H_0$ has the most impact on the Hubble constant posterior, while the exact distance value acts similarly to a normalization factor. However, future analyses that aim at precision measurements of cosmological parameters should include a more sophisticated calculation of $\beta(H_0)$, where the detectability of events following a realistic BBH mass function (rather than a Dirac delta centered on a specific mass) and the distributions of the other relevant binary parameters are taken into account.} 

We can rewrite Eq. (\ref{eq:posterior}) as:
\begin{widetext}
\begin{equation}
p(H_0|d_{\rm GW}, d_{\rm EM}) \propto \frac{p(H_0)}{V[d_{L,{\rm GW}}^{\rm max}(H_0)]} \sum_i \frac{1}{\mathcal{Z}_i}\int \de z_i \, p(d_{\rm GW}|d_L(z_i,H_0),\hat{\Omega}_i) p(d_{\rm EM}|z_i) \frac{r^2(z_i)}{H(z_i)} \, ,\label{eq:like2}
\end{equation}
\end{widetext}
where $\mathcal{Z}_i = \int p(d_{\rm EM} | z_i) r^2(z_i)/H(z_i) ~\de z_i$ are evidence terms that {correctly normalize the posterior}.
We further take into account photometric redshift systematics in the form of a bias $\Delta z$, and marginalize over it. We model the bias prior $p(\Delta z)$ with a Gaussian, where the mean and standard deviation take the same value, which is the bias evaluated from the DES Y3 validation sample presented in Section \ref{sec:des}. For simplicity of notation we remove the subscript $i$ for the redshifts inside the integral, since what really differentiates the integral for each galaxy is the different EM likelihood $p_i(d_{\rm EM}|z)$:

\begin{widetext}
\begin{equation}
p(H_0|d_{\rm GW}, d_{\rm EM}) \propto \frac{p(H_0)}{V[d_{L,{\rm GW}}^{\rm max}(H_0)]} \sum_i \frac{1}{\mathcal{Z}_i}\int \de z ~\de \Delta z \, p(d_{\rm GW}|d_L(z,H_0),\hat{\Omega}_i) p_i(d_{\rm EM}|z,\Delta z) ~p(\Delta z) \frac{r^2(z)}{H(z)} \, .\label{eq:like3}
\end{equation}
\end{widetext}

The redshift PDF, shifted in redshift by the observed bias $\Delta z$, enters in Eq. (\ref{eq:like3}) through $p_i(d_{\rm EM}|z,\Delta z)$. 

{We note that in each sky map pixel the marginal GW likelihood of Eq. (\ref{eq:like3}) takes the same value for all galaxies in that pixel, therefore the sum only becomes relevant for the EM contribution to the likelihood. That sum, which aside from the bias term is effectively the sum over the photo-$z$ PDFs, can be approximated with the $\de N/ \de z$, as mentioned in Section \ref{sec:des}. This explains why in that Section we focused on validating the redshift distribution. }

We can extend this formalism and combine data from a sample of multiple GW events if we assume that the data $\{d_{{\rm GW},i}\}$ from each event $i$ are independent of each other, and the EM data $d_{{\rm EM}}$ use a fixed galaxy catalog for all events.
{If we start again from Eq. (\ref{eq:posterior}) and add a marginalization over the redshift $z_k$ and position $\hat\Omega_k$ for each galaxy $k$ out of the $N$ galaxies in the galaxy catalog:} 
\begin{equation}
\begin{split}
p(H_0|\{d_{{\rm GW},i}\},d_{{\rm EM}}) \propto p(H_0) \prod_j p(d_{{\rm GW},j},d_{{\rm EM}}|H_0) \\ 
\propto p(H_0) p(d_{{\rm EM}}|H_0) \prod_j p(d_{{\rm GW},j}|H_0) \\
\propto p(H_0) \int \de^N z_k \de^N \hat\Omega_k
p(z_k , \hat\Omega_k) \times \\
p(d_{\rm EM} |H_0,  \{ z_k, \hat\Omega_k \} ) \left[ \prod_j p(d_{\rm GW,j} |H_0, \{ z_k, \hat\Omega_k \} ) \right] \,.
\end{split}
\end{equation}
{From this equation, the same steps followed between Eq. (\ref{eq:posterior}) and Eq. (\ref{eq:like3}) can be repeated in the same fashion in order to recover the final posterior for the combined sample.}

In our analysis we assume a flat prior on $H_0$ within [20,140] $ ~{\rm km~s^{-1}~Mpc^{-1}}$, while all the other cosmological parameters are fixed (flat $\Lambda$CDM cosmology with $\Omega_{\rm m}=0.3$ and $\Omega_\Lambda = 0.7$). The choice of this prior is mostly dictated by the choices made in previous standard siren measurements, in particular \citet*{darksiren1} and \citet{LVC_O2_StS}, in order to ensure an easier comparison of the results. Similarly to these works, the width of this prior ensures that the result is mostly informed by the LIGO/Virgo and DES data, instead of by external constraints. 

In order to avoid confirmation bias, we blind our analysis when estimating the $H_0$ posterior by randomly displacing the values of the Hubble constance by an unknown amount. We only unblind after
establishing confidence in the methodology and datasets with the blinding in place.

\section{Results and Discussion} \label{results}

\begin{figure}
    \centering
        \includegraphics[width=1\linewidth]{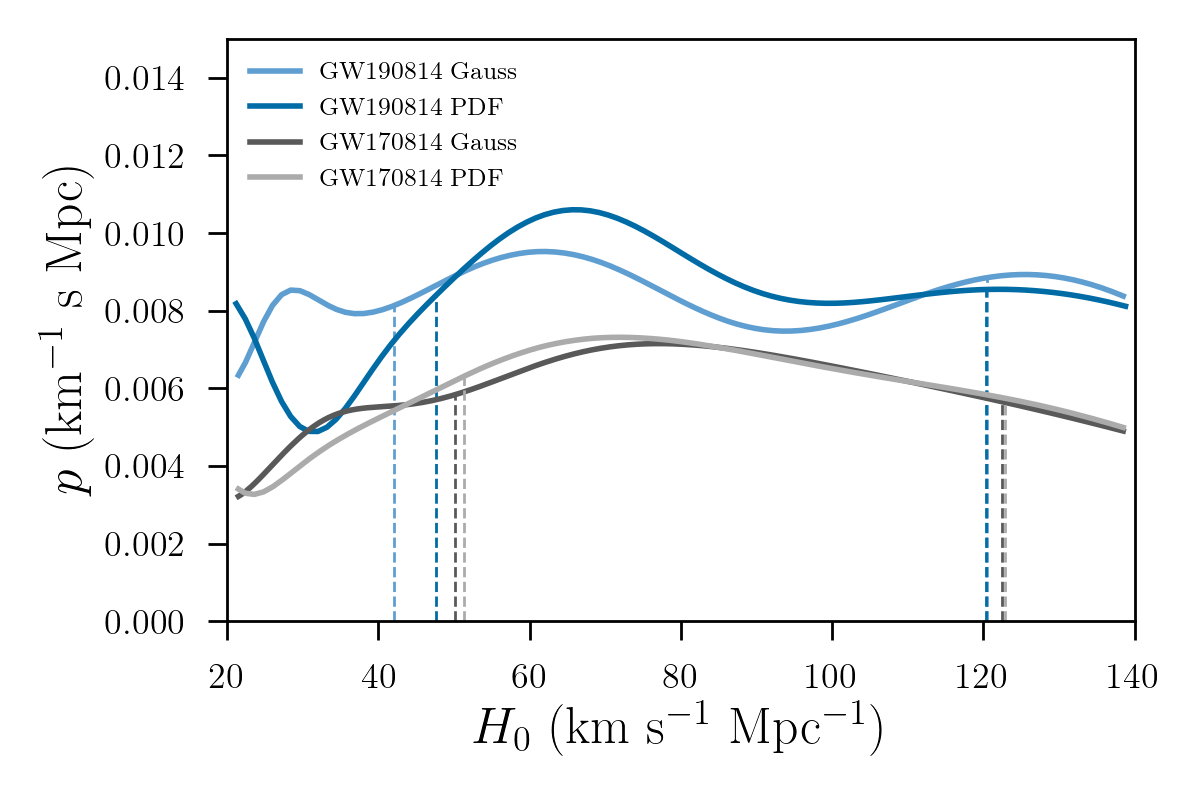}
    \caption{Hubble constant posterior distributions for GW190814 (blue curves) and GW170814 (gray curves), comparing results using the Gaussian redshift PDF versus the full PDFs. The latter case does not depend on the choice of redshift cut, as long as this contains the range of $z$ supported by the $H_0$ prior. The 68\% HDI of all PDFs is shown by the dashed lines. Posteriors have been rescaled for visualization purposes. {The sky map used here for GW190814 is the \texttt{LALInference} map released in August 2019.}}
    \label{fig:pdfs}
\end{figure}


The resulting posterior distributions for $H_0$ using GW170814 and GW190814 are shown in Figure \ref{fig:pdfs}. We contrast results for both the full redshift PDF, used in this work, and the Gaussian approximation of the redshift PDF (as done in \citealt*{darksiren1} and \citealt{LVC_O2_StS}). The difference in the shape of the posterior {is most significant at the lowest $H_0$ values}. The final 68\% {Highest Density Intervals (HDI; shown in Fig. \ref{fig:pdfs} by the dashed lines)} are consistent with each other. {It is thus unclear at this level of precision whether differences between the PDF choices are statistically significant.} However, the Gaussian approximation does not accurately reproduce a realistic photo-$z$ PDF of galaxies, which may contain multiple peaks due to color-redshift degeneracies. {Similarly, the sum of these Gaussians does not accurately reproduce the true $\de N/\de z$. Let us note that the $\de N/\de z$ is effectively what enters Eq. (7) if we approximate it with the sum of the redshift PDFs, as often done in weak lensing }(e.g. \citealt{Benjamin_2013}). On the contrary, the sum of redshift PDFs of single galaxies, or equivalently MC draws from those, are optimized to reproduce the true $\de N/\de z$ in photometric surveys such as DES and CFHTLenS (e.g. \citealt{Benjamin_2013}; \citealt*{Hoyle_2018}). Moreover, the dark siren method with the Gaussian approximated PDFs suffers from a dependence on the redshift cut applied, as was already noted in \cite*{darksiren1}. This effect is due to the contribution of the Gaussian tails, which becomes non-negligible at the high redshift end, i.e. at the high--$H_0$ end of Figure \ref{fig:pdfs}, where one marginalizes over thousands of galaxies. {Another effect of the Gaussian distributions is the flattening of the posterior, as a consequence of the overdensities being suppressed in the $\de N/\de z$.} 
The results shown here for the Gaussian approximation case correspond to a redshift cut at the $90\%$ high limit in luminosity distance, assuming the highest $H_0$ value in our prior (cf. \S\ref{sec:GWdata}). On the other hand, the posterior resulting from the use of full photo-$z$ PDFs is stable against different choices of redshift cuts (changes are at the sub--percent level). For these reasons, we choose to adopt the full PDF result as our fiducial result, and thus reanalyze GW170814 in light of these findings. For both events in our fiducial result, we adopt a conservative redshift cut corresponding to the high bound in luminosity distance at $>99.7\%$ CI for the highest $H_0$ value considered here. 

{In the future, it will be preferable to use a hierarchical, fully Bayesian framework for the galaxies' photo-z's that could be more easily incorporated into the statistical method used in this work. In fact, we notice that, while the DNF PDFs have been carefully calibrated, they do not allow a flexible implementation for Bayesian inference, and contain an ``implicit'' prior (e.g. \citealt{schmidt2020evaluation}) from the training sample that cannot be disentangled. The DES Collaboration plans to provide photo-$z$'s from methods that fit more easily within a Bayesian scheme (along with a Self-Organizing Map method, \citealt{Buchs2019}, and BPZ), and we plan on implementing those in future analyses.}

We note that the GW190814 posterior shows a clear peak between 50 and 80 ${\rm km~s}^{-1}{\rm Mpc}^{-1} $ for the PDF case, corresponding to the large overdensity shown in the top right panel of Figure \ref{fig:data}. On the other hand, both posteriors for GW170814 appear to be flatter. One main difference between the two events is that GW170814 has a significantly larger localization volume, i.e. we marginalize over a larger number of galaxies ($\sim 77,000$ versus $\sim 2,700$). Posteriors for events that are not well localized are more likely to provide flatter, less informative posteriors because the overdensities are more likely to be washed out in the marginalization over the galaxies, regardless of the photo--$z$ estimator chosen.

The marginalization over the \phz bias does not have a significant effect at the current level of precision, confirming our expectations. In fact, a redshift bias of $\Delta z\sim 0.002$ is expected to cause a bias on $H_0$ of the level $\Delta H_0 \sim c\Delta z /d_L\sim 2.5 ~{\rm km~s}^{-1}{\rm Mpc}^{-1} $ for GW190814 at $d_L\sim 240$ Mpc. However, the \phz~ bias also depends on redshift and it could have a more complicated effect on the final $H_0$ posterior. It is thus important to take into account these biases as we start combining events.

We note that the posteriors in Figure \ref{fig:pdfs} show significant probability at the high $H_0$ end. This effect is intrinsic to the dark siren method, therefore we do not widen the prior to larger values of $H_0$. In fact, the GW analysis only provides a luminosity distance estimate, which is consistent with arbitrarily large values of $H_0$. These values will be supported in the posterior from a dark siren analysis if there are galaxies at sufficiently large redshifts, which is the case for DES galaxies that extend out to $z>1$. The advantage of using full PDFs is that one can extend the $H_0$ prior, and thus the redshift range considered, out to larger values without biasing the final result, since the shape of the $H_0$ posterior in a given interval is insensitive to the redshift cut.

\begin{figure*}
    \centering
    \includegraphics[width=0.8\linewidth]{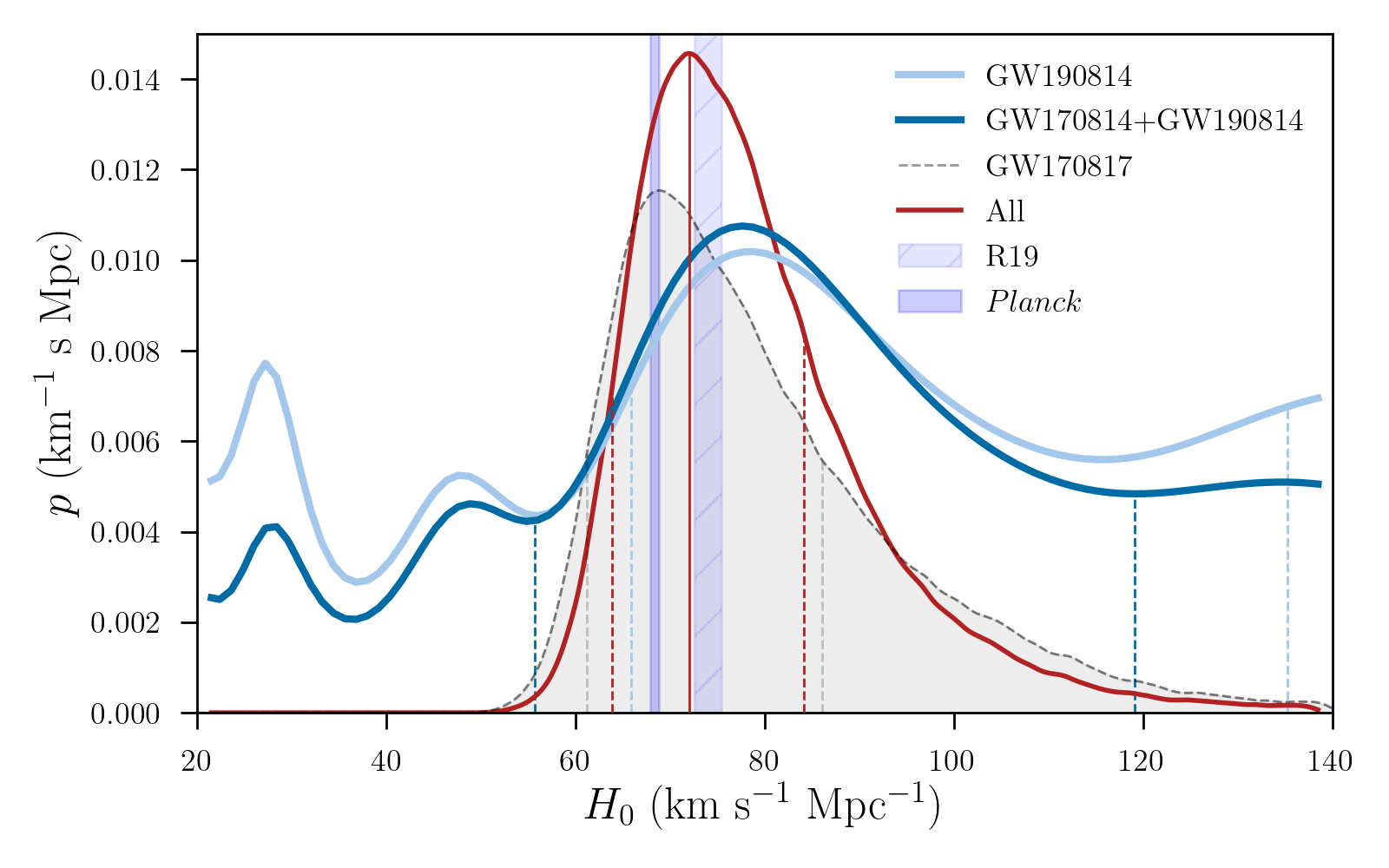}
    \caption{Hubble constant posterior distribution for GW190814 obtained by marginalizing over {$\sim 1,800$} possible host galaxies from DES (light blue line). The dark blue curve represents the posterior obtained by the combination of GW190814 and GW170814 using DES galaxies. The maximum \emph{a posteriori} and its 68\% CI for the combined result is {$H_0 = 78^{+57}_{-13}~{\rm km~s^{-1}~Mpc^{-1}}$} using a flat prior in the range [20,140] ${\rm km~s^{-1}~Mpc^{-1}}$. The posterior obtained by \citet{nicolaou2019impact} for the bright standard siren event GW170817, associated to one galaxy, is shown in gray. A combination of the two dark sirens considered in this work with GW170817 is shown in red and gives {$H_0 = 72.0^{+ 12}_{- 8.2 }~{\rm km~s^{-1}~Mpc^{-1}}$}. The addition of the dark sirens provides a $\sim 18\%$ improvement to the $68\%$ CI from GW170817 alone. The maximum \emph{a posteriori} is represented by the solid vertical line. 
    Posteriors have been rescaled for visualization purposes. The 68\% CI of all PDFs is shown by the dashed lines. {The GW190814 sky map used for this plot is the latest map released by the LVC in July 2020.}
    Constraints from \emph{Planck} (\citealt{planck18}) and \citet{Riess_2019} (R19) at 1$\sigma$ are shown in purple boxes.}
    \label{fig:result}
\end{figure*}

{Using the PDF stacking method, we then provide an $H_0$ posterior for GW190814 using the higher modes map. This result is shown in light blue in Figure \ref{fig:result}. The mode and 68\% CI of the distribution result in $H_0 = 78 ^{+ 57} _{- 13 }~{\rm km~s^{-1}~Mpc^{-1}}$. The differences between this posterior and the dark blue curve in Figure \ref{fig:pdfs}, computed with the \texttt{LALInference} map, can be explained as follows. First, the mean value of the luminosity distance marginalized over the whole sky is shifted towards lower values by $\Delta d_L=26$ Mpc in the latest map, and this results in a shift of the $H_0$ estimate of the order $\sim cz(\Delta d_L)/d_L^2\sim 8 ~{\rm km~s^{-1}~Mpc^{-1}}$ towards larger values. Secondly, the uncertainty on the luminosity distance in the \texttt{LALInference} map is almost double that in the latest map, and a larger distance uncertainty has the effect of smoothing out peaks in the $H_0$ posterior, as shown with simulations in \citep*{darksiren1}. We note that the higher modes map provides more precise distance and localization measurements \citep{190814_paper}, and thus a more precise $H_0$ measurement. Therefore, we adopt this result as our final constraint.}

{Our estimate on $H_0$ from GW190814 is consistent with those from subsection 6.4 of \citet{190814_paper}, that appeared shortly before submission of this paper. That study uses an independent pipeline and a smaller galaxy sample of $\sim 472$ galaxies (in the 90\% CI) from the GLADE catalog \citep{glade}, which is mostly incomplete at the distance of GW190814. Their analysis also finds an overdensity of galaxies around the same redshift as in this work, resulting in an $H_0$ posterior ($H_0 = 75 ^{+ 59} _{- 13}~{\rm km~s^{-1}~Mpc^{-1}}$) fully consistent with our results.}

The final posterior distribution resulting from combining the dark sirens GW190814 with GW170814 is shown in dark blue in Figure \ref{fig:result}. 
The maximum \emph{a posteriori} and the 68\% HDI around it is {$H_0 = 77 ^{+ 41} _{- 22 }~{\rm km~s^{-1}~Mpc^{-1}}$} for the combination of both dark sirens (GW190814 with GW170814), using a flat prior in the range [20,140] ${\rm km~s^{-1}~Mpc^{-1}}$ in all cases.\footnote{Note that the 68\% HDI for these multi--modal distributions is composed of two disjoint intervals. In this case we force the interval to be around the mode in one contiguous interval. The 68\% equal--tailed credible interval (i.e. the 16th and 84th quantiles) has a similar width, but we quote the HDI for consistency with previous results.}
We further combine these two dark siren events with the bright standard siren analysis of GW170817 from \citet{nicolaou2019impact}, which includes a more careful treatment of the peculiar velocity for NGC 4993 compared to \citet{2017Natur.551...85A}. We adapt the $H_0$ prior of \citet{nicolaou2019impact} to be the same used in this work. {The GW170817--only posterior is shown in Figure \ref{fig:result} by the gray shaded region, while the combined result from GW170817 with the dark sirens analyzed in this work (GW170814 plus GW190814) is given by the red solid line. We find that the addition of these dark sirens brings an {$\sim 18\%$} improvement to the 68\% HDI from GW170817 alone. For reference, we show constraints from \emph{Planck} (\citealt{planck18}) and \citet{Riess_2019} (R19) at 1$\sigma$. Our results are broadly consistent with both estimates, which is expected given the large uncertainty of the standard siren measurements. In the future, this comparison will become more interesting as we combine hundreds of events.} 

The final $H_0$ results from this and other standard siren analyses are summarized in Table \ref{tab:results}. As expected, GW170817 is the most constraining event, followed by GW190814 and GW170814, as a result of the increasing localization volume. Table \ref{tab:results} also shows the value of the 68\% HDI from the posterior divided by the 68\% CI from the flat prior, labeled as $\sigma_{H_0}/\sigma_{\rm prior}$.

In this work we have ignored the effect of peculiar velocities on the $H_0$ inference. While peculiar velocities can have a significant impact on the $H_0$ inference for nearby events, such as GW170817 (\citealt{howlett_davis,Mukherjee,palmese_WP,nicolaou2019impact}), this effect is expected to be negligible at the distance of GW190814 at the current level of precision. For example, a pessimistic bias of $300~{\rm km}~{\rm s}^{-1}$ would result in a $\sim 1 ~{\rm km~s^{-1}~Mpc^{-1}}$ bias on $H_0$. Moreover, typical biases and uncertainties on peculiar velocities are not significant compared to typical photo--$z$ uncertainties. It is however possible that at the lowest redshifts the peculiar motion is correlated across large portions of the sky, and that this effect will become relevant after we combine many events. Thus, we will address this issue in a separate work.

In this analysis we have not taken into account possible systematics introduced by the Gaussian approximation made in the sky map for the distance posterior (which enters in Eq. (\ref{eq:dpdvmap})). As posterior samples become available for more dark siren events, we will be able to assess the extent of this issue in the future. 

{It is interesting to note that both GW events overlapping with the DES footprint analyzed thus far are correlated with large overdensities. In fact, the luminosity distance of GW190814 corresponds to $z\sim 0.06$ (for $H_0=70~{\rm km~s^{-1}~Mpc^{-1}} $), and it is clear from the top right panel of Figure \ref{fig:data} that an overdensity of galaxies is present at that redshift. Similarly, there is an overdensity at the location of GW170814 at $z\sim 0.14$. This suggests that GW events could follow the underlying large scale structure (LSS), possibly with a large bias factor. Future work combining more events with LSS could confirm this hypothesis and shed light on the formation channels of these binaries (e.g. \citealt{Raccanelli_2016})}.

\begin{table*}
\centering
\begin{tabular}{cccccc}
Event & Prior  & $H_0 $ & $\sigma_{H_0}/H_0$ & $\sigma_{H_0}/\sigma_{\rm prior}$ & Reference\\
\hline
GW170817 - bright & $[20,140]$ & $68.8 ^{+ 17}_{-7.6}$ & 18\% & 31\% & Adapted from \citet{nicolaou2019impact}\\
GW170817 - dark & $[10,220]$ & $76 ^{+ 48 }_{ -23}$ & 47\% & 50\% & \citealt{fishbach}\\
GW190814 & $[20,140]$ & $78^{+57}_{-13}$ & 45\% & 86\% & This work \\
GW170814  & $[20,140]$ & $77 ^{+ 41} _{- 33 }$ & 48\% & 90\% & \citealt*{darksiren1}\\
GW170814  & $[20,140]$ & $72 ^{+ 51} _{- 20 }$ & 49\% & 87\% & This work\\
GW190814+GW170814  & $[20,140]$ & $77 ^{+ 41} _{- 22 }$ & 44\% & 77\% & This work  \\ 
All & $[20,140]$ & $72.0^{+ 12}_{- 8.2}$ & 14\% & 25\% & This work \\

\end{tabular}
\caption{Hubble constant estimates from GW170817 and the dark sirens considered in this work. All $H_0$ values and errors are in $~{\rm km~s^{-1}~Mpc^{-1}}$, all priors are flat. The uncertainty from the flat prior only is derived by assuming the same $H_0$ maximum found in the analysis. Quoted uncertainties represent 68\% HDI around the maximum of the posterior. The ``$\sigma_{H_0}/\sigma_{\rm prior}$'' column shows the 68\% CI from the posterior divided by 68\% CI of the
prior width. } 
\label{tab:results}
\end{table*}



\section{Conclusions}\label{conclusions}

In this paper we present a statistical standard siren measurement using the gravitational wave event GW190814, detected by LIGO/Virgo on August 14, 2019. The redshift information used comes from the DES DNF photo--$z$'s. Our work shows the advantages of using a full redshift PDF of the potential host galaxies versus a Gaussian approximation. Full redshift PDFs provide a better representation of the $\de N/\de z$, which, together with the luminosity distance estimates, is the ultimate ingredient needed in the dark standard siren framework. We also correct for \phz biases, however, these are subdominant at the current level of precision. Peculiar motions are also not expected to provide a significant contribution to our results at the distance of GW190814. {We run our analysis both with the preliminary GW190814 sky map released shortly after the GW alert, and with the sky map from the refined LVC analysis that uses higher order spherical harmonic modes from the GW signal. Because the latter provides better constraints on the binary parameters, we choose it as our final result.}

For GW190814, the maximum \emph{a posteriori} value of $H_0$ with its 68\% CI is $H_0 = 78 ^{+ 57} _{- 13 }~{\rm km~s^{-1}~Mpc^{-1}}$, while it is $H_0 = 77 ^{+ 41} _{- 22 }~{\rm km~s^{-1}~Mpc^{-1}}$ when combined with our re-analysis of GW170814, using a flat prior in the range [20,140] ${\rm km~s^{-1}~Mpc^{-1}}$. Finally, a combination of GW190814, GW170814 and GW170817 yields $H_0 = 72.0^{+ 12}_{- 8.2 }~{\rm km~s^{-1}~Mpc^{-1}}$. The addition of GW190814 and GW170814 to GW170817 improves the 68\% CI interval by $\sim 18\%$, showing how well--localized mergers without counterparts can provide {a substantial} contribution to standard siren measurements, provided that a complete galaxy catalog is available at the location of the event.

In the future, a combination of well--localized GW events with on--going and upcoming galaxy surveys, such as DESI, 4MOST and LSST, will be able to provide competitive constraints on the Hubble constant using the dark standard siren method. {We remind the reader that this method is not expected to provide constraints that are competitive with current precision measurements of $H_0$ from single events. In fact, the Hubble constant posteriors for single events can present multiple peaks or be asymmetric distributions, as is the case in this work and in previous dark siren analyses. These effects are due to the presence of overdensities or underdensities along the line of sight at the position of the events, and those will differ from event to event. However, after enough events are combined, the analysis will converge to the true value of $H_0$ (e.g. \citealt{2020PhRvD.101l2001G}): the background or foreground overdensities will not be consistently present for the same values of $H_0$, and their contribution to the final posterior will eventually be washed out.} If we assume that the statistical uncertainty on $H_0$ scales as $1/\sqrt{N}$, where $N$ is the number of events considered, a combination of $\mathcal{O}(100)$ events with a localization similar to GW170814 and GW190814, will provide a $\mathcal{O}(1)\%$ \emph{statistical} precision on the Hubble constant. We stress that a range of systematic effects and biases that have been neglected so far will become important at that level of precision (for example, assumptions about the background cosmology; \citealt{Keeley_2019,Shafieloo_2020}), and will need to be addressed in the coming years. An event sample of this size will in fact be available a few years after LIGO/Virgo run at design sensitivity \citep{2018LRR....21....3A}.

\acknowledgments

Funding for the DES Projects has been provided by the DOE and NSF(USA), MEC/MICINN/MINECO(Spain), STFC(UK), HEFCE(UK). NCSA(UIUC), KICP(U. Chicago), CCAPP(Ohio State), 
MIFPA(Texas A\&M), CNPQ, FAPERJ, FINEP (Brazil), DFG(Germany) and the Collaborating Institutions in the Dark Energy Survey.

The Collaborating Institutions are Argonne Lab, UC Santa Cruz, University of Cambridge, CIEMAT-Madrid, University of Chicago, University College London, 
DES-Brazil Consortium, University of Edinburgh, ETH Z{\"u}rich, Fermilab, University of Illinois, ICE (IEEC-CSIC), IFAE Barcelona, Lawrence Berkeley Lab, 
LMU M{\"u}nchen and the associated Excellence Cluster Universe, University of Michigan, NOAO, University of Nottingham, Ohio State University, University of 
Pennsylvania, University of Portsmouth, SLAC National Lab, Stanford University, University of Sussex, Texas A\&M University, and the OzDES Membership Consortium.

Based in part on observations at Cerro Tololo Inter-American Observatory, National Optical Astronomy Observatory, which is operated by the Association of 
Universities for Research in Astronomy (AURA) under a cooperative agreement with the National Science Foundation.

The DES Data Management System is supported by the NSF under Grant Numbers AST-1138766 and AST-1536171. The DES participants from Spanish institutions are partially 
supported by MINECO under grants AYA2015-71825, ESP2015-88861, FPA2015-68048, and Centro de Excelencia SEV-2016-0588, SEV-2016-0597 and MDM-2015-0509. Research leading 
to these results has received funding from the ERC under the EU's 7$^{\rm th}$ Framework Programme including grants ERC 240672, 291329 and 306478.
We acknowledge support from the Australian Research Council Centre of Excellence for All-sky Astrophysics (CAASTRO), through project number CE110001020.

This manuscript has been authored by Fermi Research Alliance, LLC under Contract No. DE-AC02-07CH11359 with the U.S. Department of Energy, Office of Science, Office of High Energy Physics. The United States Government retains and the publisher, by accepting the article for publication, acknowledges that the United States Government retains a non-exclusive, paid-up, irrevocable, world-wide license to publish or reproduce the published form of this manuscript, or allow others to do so, for United States Government purposes.

AP thanks Zoheyr Doctor, Maya Fishbach and Will Farr for very useful discussion on standard sirens. In this work we made extensive use of TOPCAT \citep{topcat}.

\bibliographystyle{yahapj_twoauthor_arxiv_amp}
\bibliography{references}

\begin{thebibliography}{}
\providecommand\natexlab[1]{#1}
\providecommand\JournalTitle[1]{#1}
\providecommand{\eprint}[1][]{\url{#1}}

\bibitem[{Abbott {et~al.}(2016)Abbott \& Abbott {et~al.}}]{GW150914}
Abbott, B.~P., Abbott, R., Abbott, T.~D., {et~al.} 2016,
  \href{http://dx.doi.org/10.1103/PhysRevLett.116.061102}{\JournalTitle{Phys.
  Rev. Lett.}, 116, 061102}

\bibitem[{{Abbott} {et~al.}(2017{\natexlab{a}}){Abbott} \& {Abbott}
  {et~al.}}]{2017Natur.551...85A}
{Abbott}, B.~P., {Abbott}, R., {Abbott}, T.~D., {et~al.} 2017{\natexlab{a}},
  \href{http://dx.doi.org/10.1038/nature24471}{\JournalTitle{\nat}, 551, 85},
  \eprint arXiv:{1710.05835}

\bibitem[{{Abbott} {et~al.}(2017{\natexlab{b}}){Abbott} \& {Abbott}
  {et~al.}}]{gw170814}
{Abbott}, B.~P., {Abbott}, R., {Abbott}, T.~D., {et~al.} 2017{\natexlab{b}},
  \href{http://dx.doi.org/10.1103/PhysRevLett.119.141101}{\JournalTitle{Phys.
  Rev. Lett.}, 119, 141101}, \eprint arXiv:{1709.09660}

\bibitem[{Abbott {et~al.}(2017)Abbott \& Abbott {et~al.}}]{ligobns}
Abbott, B.~P., Abbott, R., Abbott, T.~D., {et~al.} 2017,
  \href{http://dx.doi.org/10.1103/PhysRevLett.119.161101}{\JournalTitle{Phys.
  Rev. Lett.}, 119, 161101}

\bibitem[{Abbott {et~al.}(2020)Abbott \& Abbott {et~al.}}]{190814_paper}
Abbott, R., Abbott, T.~D., Abraham, S., {et~al.} 2020,
  \href{http://dx.doi.org/10.3847/2041-8213/ab960f}{\JournalTitle{\apj}, 896,
  L44}

\bibitem[{Ackley {et~al.}(2020)Ackley \& Amati {et~al.}}]{Ackley20}
Ackley, K., Amati, L., Barbieri, C., {et~al.} 2020, \eprint arXiv:{2002.01950}

\bibitem[{{Allen}(1976)}]{allen}
{Allen}, D.~A. 1976,
  \href{http://dx.doi.org/10.1093/mnras/174.1.29P}{\JournalTitle{\mnras}, 174,
  29P}

\bibitem[{{Andreoni} {et~al.}(2020){Andreoni} \& {Goldstein} {et~al.}}]{growth}
{Andreoni}, I., {Goldstein}, D.~A., {Kasliwal}, M.~M., {et~al.} 2020,
  \href{http://dx.doi.org/10.3847/1538-4357/ab6a1b}{\JournalTitle{\apj}, 890,
  131}, \eprint arXiv:{1910.13409}

\bibitem[{{Arcavi} {et~al.}(2017){Arcavi} \& {Hosseinzadeh} {et~al.}}]{arcavi}
{Arcavi}, I., {Hosseinzadeh}, G., {Howell}, D.~A., {et~al.} 2017,
  \href{http://dx.doi.org/10.1038/nature24291}{\JournalTitle{\nat}, 551, 64},
  \eprint arXiv:{1710.05843}

\bibitem[{{Arnouts} {et~al.}(1999){Arnouts} \& {Cristiani} {et~al.}}]{arnouts}
{Arnouts}, S., {Cristiani}, S., {Moscardini}, L., {et~al.} 1999,
  \href{http://dx.doi.org/10.1046/j.1365-8711.1999.02978.x}{\JournalTitle{\mnras},
  310, 540}, \eprint{astro-ph/9902290}

\bibitem[{{Ben{\'{\i}}tez}(2000)}]{benitez}
{Ben{\'{\i}}tez}, N. 2000,
  \href{http://dx.doi.org/10.1086/308947}{\JournalTitle{\apj}, 536, 571},
  \eprint{astro-ph/9811189}

\bibitem[{Benjamin {et~al.}(2013)Benjamin \& Van~Waerbeke
  {et~al.}}]{Benjamin_2013}
Benjamin, J., Van~Waerbeke, L., Heymans, C., {et~al.} 2013,
  \href{http://dx.doi.org/10.1093/mnras/stt276}{\JournalTitle{\mnras}, 431,
  1547–1564}

\bibitem[{{Bertin} \& {Arnouts}(1996)}]{sextractor}
{Bertin}, E. \& {Arnouts}, S. 1996,
  \href{http://dx.doi.org/10.1051/aas:1996164}{\JournalTitle{Astronomy and
  Astrophysics Supplement}, 117, 393}

\bibitem[{{Bruzual} \& {Charlot}(2003)}]{bc03}
{Bruzual}, G. \& {Charlot}, S. 2003,
  \href{http://dx.doi.org/10.1046/j.1365-8711.2003.06897.x}{\JournalTitle{\mnras},
  344, 1000}, \eprint{astro-ph/0309134}

\bibitem[{{Buchs} {et~al.}(2019){Buchs} \& {Davis} {et~al.}}]{Buchs2019}
{Buchs}, R., {Davis}, C., {Gruen}, D., {et~al.} 2019,
  \href{http://dx.doi.org/10.1093/mnras/stz2162}{\JournalTitle{\mnras}, 489,
  820}, \eprint arXiv:{1901.05005}

\bibitem[{{Chabrier}(2003)}]{chabrier}
{Chabrier}, G. 2003,
  \href{http://dx.doi.org/10.1086/376392}{\JournalTitle{\pasp}, 115, 763},
  \eprint{astro-ph/0304382}

\bibitem[{{Chen} {et~al.}(2018){Chen} \& {Fishbach} \& {Holz}}]{chen17}
{Chen}, H.-Y., {Fishbach}, M., \& {Holz}, D.~E. 2018,
  \href{http://dx.doi.org/10.1038/s41586-018-0606-0}{\JournalTitle{\nat}, 562,
  545}, \eprint arXiv:{1712.06531}

\bibitem[{{Colless} {et~al.}(2001){Colless} \& {Dalton}
  {et~al.}}]{2001MNRAS.328.1039C}
{Colless}, M., {Dalton}, G., {Maddox}, S., {et~al.} 2001,
  \href{http://dx.doi.org/10.1046/j.1365-8711.2001.04902.x}{\JournalTitle{\mnras},
  328, 1039}, \eprint{astro-ph/0106498}

\bibitem[{Coulter {et~al.}(2017)Coulter \& Foley {et~al.}}]{Coulter1556}
Coulter, D.~A., Foley, R.~J., Kilpatrick, C.~D., {et~al.} 2017,
  \href{http://dx.doi.org/10.1126/science.aap9811}{\JournalTitle{Science}, 358,
  1556}, \eprint{http://science.sciencemag.org/content/358/6370/1556.full.pdf}

\bibitem[{{Cowperthwaite} {et~al.}(2016){Cowperthwaite} \& {Berger}
  {et~al.}}]{2016ApJ...826L..29C}
{Cowperthwaite}, P.~S., {Berger}, E., {Soares-Santos}, M., {et~al.} 2016,
  \href{http://dx.doi.org/10.3847/2041-8205/826/2/L29}{\JournalTitle{\apjl},
  826, L29}, \eprint arXiv:{1606.04538}

\bibitem[{Crocce {et~al.}(2015)Crocce \& Castander \& Gaztañaga \& Fosalba \&
  Carretero}]{Crocce_2015}
Crocce, M., Castander, F.~J., Gaztañaga, E., Fosalba, P., \& Carretero, J.
  2015, \href{http://dx.doi.org/10.1093/mnras/stv1708}{\JournalTitle{\mnras},
  453, 1513–1530}

\bibitem[{{Cunha} {et~al.}(2009){Cunha} \& {Lima} \& {Oyaizu} \& {Frieman} \&
  {Lin}}]{cunha}
{Cunha}, C.~E., {Lima}, M., {Oyaizu}, H., {Frieman}, J., \& {Lin}, H. 2009,
  \href{http://dx.doi.org/10.1111/j.1365-2966.2009.14908.x}{\JournalTitle{\mnras},
  396, 2379}, \eprint arXiv:{0810.2991}

\bibitem[{{D{\'a}lya} {et~al.}(2018){D{\'a}lya} \& {Galg{\'o}czi}
  {et~al.}}]{glade}
{D{\'a}lya}, G., {Galg{\'o}czi}, G., {Dobos}, L., {et~al.} 2018,
  \href{http://dx.doi.org/10.1093/mnras/sty1703}{\JournalTitle{\mnras}, 479,
  2374}, \eprint arXiv:{1804.05709}

\bibitem[{{De Vicente} {et~al.}(2016){De Vicente} \& {S{\'a}nchez} \&
  {Sevilla-Noarbe}}]{dnf}
{De Vicente}, J., {S{\'a}nchez}, E., \& {Sevilla-Noarbe}, I. 2016,
  \href{http://dx.doi.org/10.1093/mnras/stw857}{\JournalTitle{\mnras}, 459,
  3078}, \eprint arXiv:{1511.07623}

\bibitem[{{Del Pozzo}(2012)}]{delpozzo}
{Del Pozzo}, W. 2012,
  \href{http://dx.doi.org/10.1103/PhysRevD.86.043011}{\JournalTitle{\prd}, 86,
  043011}, \eprint arXiv:{1108.1317}

\bibitem[{{DES Collaboration} {et~al.}(2018){DES Collaboration} \& {Abbott}
  {et~al.}}]{dr1}
{DES Collaboration}, {Abbott}, T.~M.~C., {Abdalla}, F.~B., {et~al.} 2018,
  \href{http://dx.doi.org/10.3847/1538-4365/aae9f0}{\JournalTitle{\apjs}, 239,
  18}, \eprint arXiv:{1801.03181}

\bibitem[{{Doctor} {et~al.}(2019){Doctor} \& {Kessler} {et~al.}}]{doctor}
{Doctor}, Z., {Kessler}, R., {Herner}, K., {et~al.} 2019,
  \href{http://dx.doi.org/10.3847/2041-8213/ab08a3}{\JournalTitle{\apjl}, 873,
  L24}, \eprint arXiv:{1812.01579}

\bibitem[{{Etherington} {et~al.}(2017){Etherington} \& {Thomas}
  {et~al.}}]{Etherington}
{Etherington}, J., {Thomas}, D., {Maraston}, C., {et~al.} 2017,
  \href{http://dx.doi.org/10.1093/mnras/stw3069}{\JournalTitle{\mnras}, 466,
  228}, \eprint arXiv:{1701.06066}

\bibitem[{{Feeney} {et~al.}(2019){Feeney} \& {Peiris}
  {et~al.}}]{2019PhRvL.122f1105F}
{Feeney}, S.~M., {Peiris}, H.~V., {Williamson}, A.~R., {et~al.} 2019,
  \href{http://dx.doi.org/10.1103/PhysRevLett.122.061105}{\JournalTitle{\prl},
  122, 061105}, \eprint arXiv:{1802.03404}

\bibitem[{{Fishbach} {et~al.}(2019){Fishbach} \& {Gray} {et~al.}}]{fishbach}
{Fishbach}, M., {Gray}, R., {Maga{\~n}a Hernandez}, I., {et~al.} 2019,
  \href{http://dx.doi.org/10.3847/2041-8213/aaf96e}{\JournalTitle{\apjl}, 871,
  L13}, \eprint arXiv:{1807.05667}

\bibitem[{{Flaugher} {et~al.}(2015){Flaugher} \& {Diehl} {et~al.}}]{flaugher}
{Flaugher}, B., {Diehl}, H.~T., {Honscheid}, K., {et~al.} 2015,
  \href{http://dx.doi.org/10.1088/0004-6256/150/5/150}{\JournalTitle{\aj}, 150,
  150}, \eprint arXiv:{1504.02900}

\bibitem[{Fosalba {et~al.}(2015)Fosalba \& Crocce \& Gaztañaga \&
  Castander}]{Fosalba_2015}
Fosalba, P., Crocce, M., Gaztañaga, E., \& Castander, F.~J. 2015,
  \href{http://dx.doi.org/10.1093/mnras/stv138}{\JournalTitle{\mnras}, 448,
  2987–3000}

\bibitem[{{Freedman}(2017)}]{2017NatAs...1E.169F}
{Freedman}, W.~L. 2017,
  \href{http://dx.doi.org/10.1038/s41550-017-0169}{\JournalTitle{Nature
  Astronomy}, 1, 0169}, \eprint arXiv:{1706.02739}

\bibitem[{Freedman {et~al.}(2019)Freedman \& Madore {et~al.}}]{Freedman_2019}
Freedman, W.~L., Madore, B.~F., Hatt, D., {et~al.} 2019,
  \href{http://dx.doi.org/10.3847/1538-4357/ab2f73}{\JournalTitle{\apj}, 882,
  34}

\bibitem[{Garcia {et~al.}(2020)}]{Garcia:2020smy}
Garcia, A. {et~al.} 2020, \eprint arXiv:{2007.00050}

\bibitem[{Gomez {et~al.}(2019)Gomez \& Hosseinzadeh {et~al.}}]{Gomez20}
Gomez, S., Hosseinzadeh, G., Cowperthwaite, P.~S., {et~al.} 2019,
  \href{http://dx.doi.org/10.3847/2041-8213/ab4ad5}{\JournalTitle{\apj}, 884,
  L55}

\bibitem[{{G{\'o}rski} {et~al.}(2005){G{\'o}rski} \& {Hivon}
  {et~al.}}]{healpix}
{G{\'o}rski}, K.~M., {Hivon}, E., {Banday}, A.~J., {et~al.} 2005,
  \href{http://dx.doi.org/10.1086/427976}{\JournalTitle{\apj}, 622, 759},
  \eprint{astro-ph/0409513}

\bibitem[{Gray {et~al.}(2019)Gray \& Hernandez {et~al.}}]{gray2019cosmological}
Gray, R., Hernandez, I.~M., Qi, H., {et~al.} 2019, Cosmological Inference using
  Gravitational Wave Standard Sirens: A Mock Data Challenge, \eprint
  arXiv:{1908.06050}

\bibitem[{{Gray} {et~al.}(2020){Gray} \& {Hernandez}
  {et~al.}}]{2020PhRvD.101l2001G}
{Gray}, R., {Hernandez}, I.~M., {Qi}, H., {et~al.} 2020,
  \href{http://dx.doi.org/10.1103/PhysRevD.101.122001}{\JournalTitle{\prd},
  101, 122001}, \eprint arXiv:{1908.06050}

\bibitem[{{Gschwend} {et~al.}(2018){Gschwend} \& {Rossel} {et~al.}}]{julia}
{Gschwend}, J., {Rossel}, A.~C., {Ogando}, R.~L.~C., {et~al.} 2018,
  \href{http://dx.doi.org/10.1016/j.ascom.2018.08.008}{\JournalTitle{Astronomy
  and Computing}, 25, 58}, \eprint arXiv:{1708.05643}

\bibitem[{{Hartley} {et~al.}(2013){Hartley} \& {Almaini} {et~al.}}]{Hartley13}
{Hartley}, W.~G., {Almaini}, O., {Mortlock}, A., {et~al.} 2013,
  \href{http://dx.doi.org/10.1093/mnras/stt383}{\JournalTitle{\mnras}, 431,
  3045}, \eprint arXiv:{1303.0816}

\bibitem[{{Holz} \& {Hughes}(2005)}]{2005ApJ...629...15H}
{Holz}, D.~E. \& {Hughes}, S.~A. 2005,
  \href{http://dx.doi.org/10.1086/431341}{\JournalTitle{\apj}, 629, 15},
  \eprint{astro-ph/0504616}

\bibitem[{{Howlett} \& {Davis}(2020)}]{howlett_davis}
{Howlett}, C. \& {Davis}, T.~M. 2020,
  \href{http://dx.doi.org/10.1093/mnras/staa049}{\JournalTitle{\mnras}, 492,
  3803}, \eprint arXiv:{1909.00587}

\bibitem[{Hoyle {et~al.}(2018)Hoyle \& Gruen {et~al.}}]{Hoyle_2018}
Hoyle, B., Gruen, D., Bernstein, G.~M., {et~al.} 2018,
  \href{http://dx.doi.org/10.1093/mnras/sty957}{\JournalTitle{\mnras}, 478,
  592–610}

\bibitem[{{Ilbert} {et~al.}(2006){Ilbert} \& {Arnouts}
  {et~al.}}]{ilbertlephare}
{Ilbert}, O., {Arnouts}, S., {McCracken}, H.~J., {et~al.} 2006,
  \href{http://dx.doi.org/10.1051/0004-6361:20065138}{\JournalTitle{\aap}, 457,
  841}, \eprint{astro-ph/0603217}

\bibitem[{{Jones} {et~al.}(2009){Jones} \& {Read}
  {et~al.}}]{2009MNRAS.399..683J}
{Jones}, D.~H., {Read}, M.~A., {Saunders}, W., {et~al.} 2009,
  \href{http://dx.doi.org/10.1111/j.1365-2966.2009.15338.x}{\JournalTitle{\mnras},
  399, 683}, \eprint arXiv:{0903.5451}

\bibitem[{{KAGRA Collaboration} {et~al.}(2018){KAGRA Collaboration} \& {LIGO
  Scientific Collaboration} \& {Virgo Collaboration}}]{2018LRR....21....3A}
{KAGRA Collaboration}, {LIGO Scientific Collaboration}, \& {Virgo
  Collaboration}. 2018,
  \href{http://dx.doi.org/10.1007/s41114-018-0012-9}{\JournalTitle{Living
  Reviews in Relativity}, 21, 3}, \eprint arXiv:{1304.0670}

\bibitem[{Keeley {et~al.}(2019)Keeley \& Shafieloo \& L’Huillier \&
  Linder}]{Keeley_2019}
Keeley, R.~E., Shafieloo, A., L’Huillier, B., \& Linder, E.~V. 2019,
  \href{http://dx.doi.org/10.1093/mnras/stz3304}{\JournalTitle{\mnras}, 491,
  3983–3989}

\bibitem[{{LIGO Scientific Collaboration} \& {Virgo
  Collaboration}(2019{\natexlab{a}})}]{gcn_skymap}
{LIGO Scientific Collaboration} \& {Virgo Collaboration}. 2019{\natexlab{a}},
  \JournalTitle{GCN 25333}

\bibitem[{{LIGO Scientific Collaboration} \& {Virgo
  Collaboration}(2019{\natexlab{b}})}]{LVC_O2_StS}
{LIGO Scientific Collaboration} \& {Virgo Collaboration}. 2019{\natexlab{b}}, A
  gravitational-wave measurement of the Hubble constant following the second
  observing run of Advanced LIGO and Virgo, \eprint arXiv:{1908.06060}

\bibitem[{{LIGO Scientific Collaboration} {et~al.}(2017){LIGO Scientific
  Collaboration} \& {Virgo Collaboration} {et~al.}}]{MMApaper}
{LIGO Scientific Collaboration}, {Virgo Collaboration}, {GBM}, F., {et~al.}
  2017, \href{http://dx.doi.org/10.3847/2041-8213/aa91c9}{\JournalTitle{\apjl},
  848, L12}, \eprint arXiv:{1710.05833}

\bibitem[{{Lima} {et~al.}(2008){Lima} \& {Cunha} \& {Oyaizu} \& {Frieman} \&
  {Lin} \& {Sheldon}}]{lima}
{Lima}, M., {Cunha}, C.~E., {Oyaizu}, H., {et~al.} 2008,
  \href{http://dx.doi.org/10.1111/j.1365-2966.2008.13510.x}{\JournalTitle{\mnras},
  390, 118}, \eprint arXiv:{0801.3822}

\bibitem[{{Lipunov} {et~al.}(2017){Lipunov} \& {Gorbovskoy} {et~al.}}]{lipunov}
{Lipunov}, V.~M., {Gorbovskoy}, E., {Kornilov}, V.~G., {et~al.} 2017,
  \href{http://dx.doi.org/10.3847/2041-8213/aa92c0}{\JournalTitle{\apjl}, 850,
  L1}, \eprint arXiv:{1710.05461}

\bibitem[{{MacLeod} \& {Hogan}(2008)}]{macleod}
{MacLeod}, C.~L. \& {Hogan}, C.~J. 2008,
  \href{http://dx.doi.org/10.1103/PhysRevD.77.043512}{\JournalTitle{\prd}, 77,
  043512}, \eprint arXiv:{0712.0618}

\bibitem[{{Mandel} {et~al.}(2019){Mandel} \& {Farr} \& {Gair}}]{mandel}
{Mandel}, I., {Farr}, W.~M., \& {Gair}, J.~R. 2019,
  \href{http://dx.doi.org/10.1093/mnras/stz896}{\JournalTitle{\mnras}, 486,
  1086}, \eprint arXiv:{1809.02063}

\bibitem[{Morgan {et~al.}(2020)Morgan \& Soares-Santos
  {et~al.}}]{morgan2020constraints}
Morgan, R., Soares-Santos, M., Annis, J., {et~al.} 2020, Constraints on the
  Physical Properties of S190814bv through Simulations based on DECam Follow-up
  Observations by the Dark Energy Survey, \eprint arXiv:{2006.07385}

\bibitem[{{Morganson} {et~al.}(2018){Morganson} \& {Gruendl}
  {et~al.}}]{2018PASP..130g4501M}
{Morganson}, E., {Gruendl}, R.~A., {Menanteau}, F., {et~al.} 2018,
  \href{http://dx.doi.org/10.1088/1538-3873/aab4ef}{\JournalTitle{\pasp}, 130,
  074501}, \eprint arXiv:{1801.03177}

\bibitem[{{Mortlock} {et~al.}(2019){Mortlock} \& {Feeney} \& {Peiris} \&
  {Williamson} \& {Nissanke}}]{2019PhRvD.100j3523M}
{Mortlock}, D.~J., {Feeney}, S.~M., {Peiris}, H.~V., {Williamson}, A.~R., \&
  {Nissanke}, S.~M. 2019,
  \href{http://dx.doi.org/10.1103/PhysRevD.100.103523}{\JournalTitle{\prd},
  100, 103523}, \eprint arXiv:{1811.11723}

\bibitem[{{M{\"o}rtsell} \& {Dhawan}(2018)}]{2018JCAP...09..025M}
{M{\"o}rtsell}, E. \& {Dhawan}, S. 2018,
  \href{http://dx.doi.org/10.1088/1475-7516/2018/09/025}{\JournalTitle{\jcap},
  9, 025}, \eprint arXiv:{1801.07260}

\bibitem[{{Mukherjee} {et~al.}(2019){Mukherjee} \& {Lavaux}
  {et~al.}}]{Mukherjee}
{Mukherjee}, S., {Lavaux}, G., {Bouchet}, F.~R., {et~al.} 2019,
  \JournalTitle{arXiv e-prints}, arXiv:1909.08627, \eprint arXiv:{1909.08627}

\bibitem[{{Nair} {et~al.}(2018){Nair} \& {Bose} \& {Saini}}]{nair}
{Nair}, R., {Bose}, S., \& {Saini}, T.~D. 2018,
  \href{http://dx.doi.org/10.1103/PhysRevD.98.023502}{\JournalTitle{\prd}, 98,
  023502}, \eprint arXiv:{1804.06085}

\bibitem[{Nicolaou {et~al.}(2020)Nicolaou \& Lahav \& Lemos \& Hartley \&
  Braden}]{nicolaou2019impact}
Nicolaou, C., Lahav, O., Lemos, P., Hartley, W., \& Braden, J. 2020,
  \href{http://dx.doi.org/10.1093/mnras/staa1120}{\JournalTitle{\mnras}, 495,
  90}, \eprint arXiv:{1909.09609}

\bibitem[{{Nishizawa}(2017)}]{2017PhRvD..96j1303N}
{Nishizawa}, A. 2017,
  \href{http://dx.doi.org/10.1103/PhysRevD.96.101303}{\JournalTitle{\prd}, 96,
  101303}, \eprint arXiv:{1612.06060}

\bibitem[{{Nissanke} {et~al.}(2013){Nissanke} \& {Holz} \& {Dalal} \& {Hughes}
  \& {Sievers} \& {Hirata}}]{2013arXiv1307.2638N}
{Nissanke}, S., {Holz}, D.~E., {Dalal}, N., {et~al.} 2013, \JournalTitle{ArXiv
  e-prints}, \eprint arXiv:{1307.2638}

\bibitem[{{Nissanke} {et~al.}(2010){Nissanke} \& {Holz} \& {Hughes} \& {Dalal}
  \& {Sievers}}]{2010ApJ...725..496N}
{Nissanke}, S., {Holz}, D.~E., {Hughes}, S.~A., {Dalal}, N., \& {Sievers},
  J.~L. 2010,
  \href{http://dx.doi.org/10.1088/0004-637X/725/1/496}{\JournalTitle{\apj},
  725, 496}, \eprint arXiv:{0904.1017}

\bibitem[{{Oyaizu} {et~al.}(2008){Oyaizu} \& {Lima} \& {Cunha} \& {Lin} \&
  {Frieman}}]{oyaizu}
{Oyaizu}, H., {Lima}, M., {Cunha}, C.~E., {Lin}, H., \& {Frieman}, J. 2008,
  \href{http://dx.doi.org/10.1086/592591}{\JournalTitle{\apj}, 689, 709},
  \eprint arXiv:{0711.0962}

\bibitem[{{Palmese} \& {Kim}(2020)}]{palmese20}
{Palmese}, A. \& {Kim}, A.~G. 2020, \JournalTitle{arXiv e-prints}, \eprint
  arXiv:{2005.04325}

\bibitem[{{Palmese} {et~al.}(2016){Palmese} \& {Lahav} {et~al.}}]{palmese16}
{Palmese}, A., {Lahav}, O., {Banerji}, M., {et~al.} 2016,
  \href{http://dx.doi.org/10.1093/mnras/stw2062}{\JournalTitle{\mnras}, 463,
  1486}, \eprint arXiv:{1601.00589}

\bibitem[{{Palmese} {et~al.}(2017){Palmese} \& {Hartley} {et~al.}}]{palmese}
{Palmese}, A., {Hartley}, W., {Tarsitano}, F., {et~al.} 2017,
  \href{http://dx.doi.org/10.3847/2041-8213/aa9660}{\JournalTitle{\apjl}, 849,
  L34}, \eprint arXiv:{1710.06748}

\bibitem[{{Palmese} {et~al.}(2019){Palmese} \& {Graur} {et~al.}}]{palmese_WP}
{Palmese}, A., {Graur}, O., {Annis}, J.~T., {et~al.} 2019,
  \JournalTitle{\baas}, 51, 310, \eprint arXiv:{1903.04730}

\bibitem[{{Palmese} {et~al.}(2020){Palmese} \& {Annis} {et~al.}}]{palmese18}
{Palmese}, A., {Annis}, J., {Burgad}, J., {et~al.} 2020,
  \href{http://dx.doi.org/10.1093/mnras/staa526}{\JournalTitle{\mnras}, 493,
  4591}, \eprint arXiv:{1903.08813}

\bibitem[{{Planck Collaboration} {et~al.}(2018){Planck Collaboration} \&
  {Aghanim} {et~al.}}]{planck18}
{Planck Collaboration}, {Aghanim}, N., {Akrami}, Y., {et~al.} 2018,
  \JournalTitle{ArXiv e-prints}, \eprint arXiv:{1807.06209}

\bibitem[{{Pozzetti} {et~al.}(2010){Pozzetti} \& {Bolzonella}
  {et~al.}}]{Pozzetti10}
{Pozzetti}, L., {Bolzonella}, M., {Zucca}, E., {et~al.} 2010,
  \href{http://dx.doi.org/10.1051/0004-6361/200913020}{\JournalTitle{\aap},
  523, A13}, \eprint arXiv:{0907.5416}

\bibitem[{Raccanelli {et~al.}(2016)Raccanelli \& Kovetz \& Bird \& Cholis \&
  Muñoz}]{Raccanelli_2016}
Raccanelli, A., Kovetz, E.~D., Bird, S., Cholis, I., \& Muñoz, J.~B. 2016,
  \href{http://dx.doi.org/10.1103/physrevd.94.023516}{\JournalTitle{\prd}, 94}

\bibitem[{Riess {et~al.}(2019)Riess \& Casertano \& Yuan \& Macri \&
  Scolnic}]{Riess_2019}
Riess, A.~G., Casertano, S., Yuan, W., Macri, L.~M., \& Scolnic, D. 2019,
  \href{http://dx.doi.org/10.3847/1538-4357/ab1422}{\JournalTitle{\apj}, 876,
  85}

\bibitem[{Schmidt {et~al.}(2020)Schmidt \& Malz
  {et~al.}}]{schmidt2020evaluation}
Schmidt, S.~J., Malz, A.~I., Soo, J. Y.~H., {et~al.} 2020, Evaluation of
  probabilistic photometric redshift estimation approaches for LSST, \eprint
  arXiv:{2001.03621}

\bibitem[{{Schutz}(1986)}]{schutz}
{Schutz}, B.~F. 1986,
  \href{http://dx.doi.org/10.1038/323310a0}{\JournalTitle{\nat}, 323, 310}

\bibitem[{Shafieloo {et~al.}(2020)Shafieloo \& Keeley \&
  Linder}]{Shafieloo_2020}
Shafieloo, A., Keeley, R.~E., \& Linder, E.~V. 2020,
  \href{http://dx.doi.org/10.1088/1475-7516/2020/03/019}{\JournalTitle{Journal
  of Cosmology and Astroparticle Physics}, 2020, 019–019}

\bibitem[{{Singer} {et~al.}(2016){Singer} \& {Chen} {et~al.}}]{Singer}
{Singer}, L.~P., {Chen}, H.-Y., {Holz}, D.~E., {et~al.} 2016,
  \href{http://dx.doi.org/10.3847/0067-0049/226/1/10}{\JournalTitle{\apjs},
  226, 10}, \eprint arXiv:{1605.04242}

\bibitem[{{Soares-Santos} {et~al.}(2016){Soares-Santos} \& {Kessler}
  {et~al.}}]{2016ApJ...823L..33S}
{Soares-Santos}, M., {Kessler}, R., {Berger}, E., {et~al.} 2016,
  \href{http://dx.doi.org/10.3847/2041-8205/823/2/L33}{\JournalTitle{\apjl},
  823, L33}, \eprint arXiv:{1602.04198}

\bibitem[{{Soares-Santos} {et~al.}(2017){Soares-Santos} \& {Holz}
  {et~al.}}]{marcelle17}
{Soares-Santos}, M., {Holz}, D.~E., {Annis}, J., {et~al.} 2017,
  \href{http://dx.doi.org/10.3847/2041-8213/aa9059}{\JournalTitle{\apjl}, 848,
  L16}, \eprint arXiv:{1710.05459}

\bibitem[{{Soares-Santos} {et~al.}(2019){Soares-Santos} \& {Palmese}
  {et~al.}}]{darksiren1}
{Soares-Santos}, M., {Palmese}, A., {Hartley}, W., {et~al.} 2019,
  \href{http://dx.doi.org/10.3847/2041-8213/ab14f1}{\JournalTitle{\apjl}, 876,
  L7}, \eprint arXiv:{1901.01540}

\bibitem[{{Tanvir} {et~al.}(2017){Tanvir} \& {Levan} {et~al.}}]{tanvir}
{Tanvir}, N.~R., {Levan}, A.~J., {Gonz{\'a}lez-Fern{\'a}ndez}, C., {et~al.}
  2017, \href{http://dx.doi.org/10.3847/2041-8213/aa90b6}{\JournalTitle{\apjl},
  848, L27}, \eprint arXiv:{1710.05455}

\bibitem[{{Taylor}(2005)}]{topcat}
{Taylor}, M.~B. 2005, Astronomical Society of the Pacific Conference Series,
  Vol. 347, {TOPCAT \&amp; STIL: Starlink Table/VOTable Processing Software},
  ed. P.~{Shopbell}, M.~{Britton}, \& R.~{Ebert}, 29

\bibitem[{{The Dark Energy Survey Collaboration}(2005)}]{descollaboration}
{The Dark Energy Survey Collaboration}. 2005, \JournalTitle{preprint
  (arXiv:astro-ph/0510346)}, \eprint{astro-ph/0510346}

\bibitem[{{The Dark Energy Survey Collaboration}(2016)}]{2016MNRAS.460.1270D}
{The Dark Energy Survey Collaboration}. 2016,
  \href{http://dx.doi.org/10.1093/mnras/stw641}{\JournalTitle{\mnras}, 460,
  1270}, \eprint arXiv:{1601.00329}

\bibitem[{{Valenti} {et~al.}(2017){Valenti} \& {David} {et~al.}}]{valenti}
{Valenti}, S., {David}, {Sand}, J., {et~al.} 2017,
  \href{http://dx.doi.org/10.3847/2041-8213/aa8edf}{\JournalTitle{\apjl}, 848,
  L24}, \eprint arXiv:{1710.05854}

\bibitem[{{Verde} {et~al.}(2019){Verde} \& {Treu} \& {Riess}}]{verde}
{Verde}, L., {Treu}, T., \& {Riess}, A.~G. 2019,
  \href{http://dx.doi.org/10.1038/s41550-019-0902-0}{\JournalTitle{Nature
  Astronomy}, 3, 891}, \eprint arXiv:{1907.10625}

\bibitem[{{Vieira} {et~al.}(2020){Vieira} \& {Ruan} {et~al.}}]{vieira}
{Vieira}, N., {Ruan}, J.~J., {Haggard}, D., {et~al.} 2020,
  \href{http://dx.doi.org/10.3847/1538-4357/ab917d}{\JournalTitle{\apj}, 895,
  96}, \eprint arXiv:{2003.09437}

\bibitem[{{Vitale} \& {Chen}(2018)}]{2018PhRvL.121b1303V}
{Vitale}, S. \& {Chen}, H.-Y. 2018,
  \href{http://dx.doi.org/10.1103/PhysRevLett.121.021303}{\JournalTitle{Phys.
  Rev. Lett.}, 121, 021303}, \eprint arXiv:{1804.07337}

\bibitem[{Watson {et~al.}(2020)Watson \& Butler {et~al.}}]{Watson20}
Watson, A.~M., Butler, N.~R., Lee, W.~H., {et~al.} 2020,
  \href{http://dx.doi.org/10.1093/mnras/staa161}{\JournalTitle{\mnras}, 492,
  5916}

\bibitem[{{Weigel} {et~al.}(2016){Weigel} \& {Schawinski} \&
  {Bruderer}}]{weigel}
{Weigel}, A.~K., {Schawinski}, K., \& {Bruderer}, C. 2016,
  \href{http://dx.doi.org/10.1093/mnras/stw756}{\JournalTitle{\mnras}, 459,
  2150}, \eprint arXiv:{1604.00008}

\bibitem[{Yu {et~al.}(2020)Yu \& Wang \& Zhao \& Lu}]{yu2020hunting}
Yu, J., Wang, Y., Zhao, W., \& Lu, Y. 2020, Hunting for the host galaxy groups
  of binary black holes and the application in constraining Hubble constant,
  \eprint arXiv:{2003.06586}

\end{thebibliography}
\end{document}